%% file: main.tex
\documentclass[pre, aps, showpacs, superscriptaddress, twocolumn, longbibliography, groupedaddress]{revtex4-2}

\usepackage{hyperref}
\hypersetup{
     colorlinks=true,
     linkcolor=magenta,
     filecolor=blue,
     citecolor=blue,      
     urlcolor=cyan,
     }
\usepackage{color}
\usepackage[usenames,dvipsnames]{xcolor}
\usepackage{amsmath,amsthm,amssymb}
\usepackage{algorithm2e}
\RestyleAlgo{ruled}
\usepackage{graphicx}
\usepackage{float}
\usepackage{epsfig}
\usepackage{bm}% bold math
\usepackage{mathrsfs}
\usepackage{multirow}
\usepackage[all]{xy}
\usepackage{pbox}
\usepackage{verbatim}
\usepackage{braket}
\usepackage{mathtools}
\usepackage{bm}
\usepackage{tikz}
\usepackage{xcolor}
 
\usepackage{mathtools}
\usepackage{mathtools}
\usepackage{tabstackengine}
\usepackage{enumerate} 
\usepackage{soul}  
\usepackage{mathrsfs}
\stackMath
\DeclareMathOperator{\Tr}{Tr}

\newcommand{\er}[1]{Eq.~\eqref{#1}}

\newcommand{\era}[2]{Eqs.~(\ref{#1}) and (\ref{#2})}

\newcommand{\Er}[1]{Equation~\eqref{#1}}

\newcommand{\beq}{\begin{equation}}
\newcommand{\eeq}{\end{equation}}

\newcommand{\tri}{{\raisebox{-1.4pt}{$\triangle$}}}

\begin{document}  

\title{Rejection-free quantum Monte Carlo in continuous time from transition path sampling}

\author{Luke Causer}
\author{Konstantinos Sfairopoulos}
\author{Jamie F. Mair}
\author{Juan P. Garrahan}
\affiliation{School of Physics and Astronomy, University of Nottingham, Nottingham, NG7 2RD, UK}
\affiliation{Centre for the Mathematics and Theoretical Physics of Quantum Non-Equilibrium Systems,
University of Nottingham, Nottingham, NG7 2RD, UK}

\begin{abstract}
\input{abstract}
\end{abstract}

\maketitle

\input{sections/introduction}
\input{sections/boltzmann}
\input{sections/optimal}
\input{sections/ising}
\input{sections/tpm}
\input{sections/conclusions}

\begin{acknowledgments}
We acknowledge financial support from EPSRC Grant no.\ EP/V031201/1.
LC was supported by an EPSRC Doctoral prize from the University of Nottingham.
Calculations were performed using the Sulis Tier 2 HPC platform hosted by the Scientific Computing Research Technology Platform at the University of Warwick. Sulis is funded by EPSRC Grant EP/T022108/1 and the HPC Midlands+ consortium.
We acknowledge access to the University of Nottingham Augusta HPC service. 
\end{acknowledgments}

\section*{Data and code availability}
The data shown in the figures is available at Ref.~\cite{Causer2023b_data}.
Example code used to generate the data can be found at \href{https://doi.org/10.5281/zenodo.10246029}{10.5281/zenodo.10246029}.

\appendix

\input{appendix/sampling}
\input{appendix/lds}

%\bibliography{bibliography.bib}
\bibliographystyle{apsrev4-2}

\input{bibliography.bbl}
\end{document}

%% file: abstract.tex
Continuous-time quantum Monte Carlo refers to a class of algorithms designed to sample the thermal distribution of a quantum Hamiltonian through exact expansions of the Boltzmann exponential in terms of stochastic trajectories which are periodic in imaginary time. Here, we show that for (sign-problem-free) quantum many-body systems with discrete degrees of freedom --- such as spins on a lattice --- this sampling can be done in a rejection-free manner using transition path sampling (TPS). The key idea is to converge the trajectory ensemble through updates where one individual degree of freedom is modified across all time while the remaining unaltered ones provide a time-dependent background. The ensuing single-body dynamics provides a way to generate trajectory updates exactly, allowing one to obtain the target ensemble efficiently via rejection-free TPS. 
We demonstrate our method on the transverse field Ising model in one and two dimensions, and on the quantum triangular plaquette (or Newman-Moore) model. We show that despite large autocorrelation times, our method is able to efficiently recover the respective quantum phase transition of each model. We also discuss the connection to rare event sampling in continuous-time Markov dynamics.

%% file: sections/introduction.tex
\section{Introduction}

Statistical mechanics has facilitated the study of many-body systems through Monte Carlo sampling for several decades. 
Originally developed for classical Hamiltonians, classical Monte Carlo methods allow the sampling of the Boltzmann distribution by proposing updates to a configuration together with a criterion, such as Metropolis \cite{Metropolis1953, Hastings1970} or Glauber \cite{Glauber1963}, to accept or reject the changes.
At the same time, Monte Carlo sampling can be generalised to study the Boltzmann distribution of quantum many-body Hamiltonians \cite{Sandvik2019}. However, a complication here is that the weights of the configurations require the evolution to be computed in imaginary time. 

A standard approach is to split the imaginary-time evolution of the quantum partition sum into discrete time steps, giving rise to ``trajectories'' of configurations in discretised imaginary-time, where the weightings of the trajectories can be estimated for small time steps through a Trotter-Suzuki decomposition \cite{Suzuki1977, Hirsch1982}.
This procedure can be made exact by considering a {\em continuous-time} expansion instead, in terms of a perturbative expansion of the Boltzmann exponential in the interaction picture and to all orders, yielding a sum over classical trajectories in continuous time
\cite{Beard1996, Prokofev1998, Rubtsov2005, Gull2011}.
In either of these approaches, there is added complexity in relation to the classical case, as imaginary-time trajectories have to be sampled 
by making changes to the temporal degrees of freedom in addition to the spatial ones
\footnote{
Another approach worth mentioning is the stochastic series expansion (e.g. Refs.~\cite{Handscomb1964, Sandvik1991, Sandvik1992, Syljuaasen2002, Sandvik2019}), which instead utilises the Taylor expansion of the exponential in order to obtain strings of operators.
While this does not require the consideration of imaginary time, there is still an added complexity in the length of the strings of operators.
}.

In this paper we focus on the continuous-time expansion of the quantum Boltzmann distribution through trajectory ensembles.
For systems with discrete local degrees of freedom (and in the absence of a sign problem \cite{Loh1990, Pan2022}) we show how the trajectories that define the quantum partition sum can be  sampled  through the realisation of the so-called {\em Doob dynamics} (see e.g.\ Refs.~\cite{Jack2010,Chetrite2015,Garrahan2016}), which optimally samples rare events. 
Specifically, we show how this can be achieved efficiently through exact trajectory updates which are {\em local in space but extensive in time}. In this way we can generate ensembles of trajectories using a version of transition path sampling (TPS) which is {\em rejection-free} (in the sense it does not require any acceptance criteria) \cite{Krzakala2008, Mora2012}. 

We illustrate the method by studying the transverse field Ising model (TFIM) in one and two spatial dimensions. In this case we sample many-body trajectories by updating the trajectories of individual spins, while keeping the other spins as an effective time-dependent background.
This approach is similar to the one in Ref.~\cite{Krzakala2008}, which studied the TFIM on the Bethe lattice, and to the one in Ref.~\cite{Mora2012}, which studied transition events of the classical 2D Ising model.
It also generalises the approach adopted in Ref.~\cite{Vasiloiu2020} which studied the ground state properties of the 1D TFIM via time- and space-local TPS updates.
Here, we demonstrate that this approach is capable of predicting the well known continuous phase transitions for the 1D and 2D TFIMs, despite the fact that the local updates suffer from large autocorrelation times close to criticality.
It is important to note that there exist update schemes which can overcome the difficulties of criticality, for example, cluster updates \cite{Rieger1999,Blote2002} (a generalisation of the Swendsen-Wang algorithm \cite{Swendsen1987} to continuous-time ensembles), and the Worm algorithm \cite{Huang2020}. 
Reference~\cite{Sfairopoulos2023} demonstrates the use of our approach where these alternative methods are difficult to formulate.

We also show how to use our approach with more complex spin interactions 
by means of customised local updates. 
As a concrete example, we study the quantum Newman-Moore model \cite{Yoshida2014, Devakul2019, Devakul2019b, Vasiloiu2020, Zhou2021, Sfairopoulos2023, Wiedmann2023}, also known as the quantum triangular plaquette model (QTPM). This model builds on the classical TPM \cite{Newman1999, Garrahan2000, Garrahan2002}, a 
system of spins with three-body interactions studied in the context of glasses, making it quantum by adding a transverse magnetic field. 
For the QTPM we devise local updates by simultaneously changing the trajectories of three neighbouring spins.
Our method can effectively recover the first-order quantum phase transition 
of the QTPM \cite{Yoshida2014, Devakul2019, Vasiloiu2020, Zhou2021, Sfairopoulos2023}. We further show how thermal annealing can be implemented to improve sampling close to the transition point.

The paper is structured as follows.
Section~\ref{sec: single_spin} formalises the connection between the quantum Boltzmann distribution and rare-trajectory sampling, explaining how observables can be estimated from trajectories.
Section~\ref{sec: optimal} then explains how the optimal Doob dynamics can be used to sample trajectories from the quantum Boltzmann distribution.
Section~\ref{sec:ising} demonstrates how this approach can be used to incorporate a single-spin update scheme for spin models with a simple transverse field, in analogy to Ref.~\cite{Krzakala2008}, by implementing the method for the 1D and 2D TFIM.
Section~\ref{sec:tpm} generalises the method to Hamiltonians with more complex interaction terms, using the QTPM as a concrete example.
We give our conclusions in Sec.~\ref{sec: conclusions}, where we discuss the possibility of more involved schemes, including updates which are collective in space.
Appendix~\ref{appendix:statistics} demonstrates how to use our approach for rare event sampling in classical continuous-time Markov dynamics, where we determine the dynamical large deviations (LDs) of the TFIM and QTPM as an example.

%% file: sections/boltzmann.tex
\section{Monte Carlo sampling of the Boltzmann distribution}
\label{sec: single_spin}
\subsection{The quantum partition function and continuous-time stochastic dynamics}

We consider a system with a Hamiltonian $\hat{H}$ and a discrete set of configurations that defines a basis $\{\ket{x}\}$ of its Hilbert space, $\mathscr{H}$
\footnote{Which is not to be confused with a position in real space.}.
In this basis, the Hamiltonian can be decomposed into a diagonal and a off-diagonal part, $\hat{H} = \hat{H}^{\rm c} - \hat{K}$,
\begin{gather}
    \hat{H}^{\rm c} = \sum_{x} H^c_{x} \ket{x}\bra{x},
    \label{R}
    \\
    \label{K}
    \hat{K} = \sum_{x, y\neq x} \hat{K}_{x\to y} \equiv \sum_{x, y\neq x} K_{x\to y}\ket{y}\bra{x}.
\end{gather}
We will refer to operators which act on the whole Hilbert space by symbols with a hat (e.g. $\hat{H}^{c}$), and their individual matrix elements by symbols without a hat.

The statistical properties of $\hat{H}$ at some finite temperature are characterised by the partition function,
\beq
    Z_{\beta} = \Tr \left[e^{-\beta \hat{H}}\right] = \sum_{x} \braket{x | e^{-\beta \hat{H}} | x}.
    \label{Z}
\eeq
\Er{Z} can be expressed as a sum over stochastic paths by considering its Dyson series expansion \cite{Dyson1949},
\begin{multline}
    Z_{\beta} = \sum_{M=0}^{\infty} \sum_{x_{\mkern-0.5mu 0}, \cdots, x_{\mkern-2mu M}} \int dt_{\mkern-2mu M}\dots dt_{1} 
    \bra{x_{0}}  e^{-(\beta-t_{\mkern-2mu M})\hat{H}^{\rm c}}
    \\
    \prod_{m = 1}^{M} \left[ \hat{K}_{x_{\mkern-1mu m-1}\to x_{\mkern-1mu m}} e^{-(t_{m}-t_{m-1})\hat{H}^{\rm c}}\right]
    \ket{x_{0}},
    \label{Z_dyson}
\end{multline}
where the integrals over times $t_{m}$ are performed with the limits $\beta \geq t_{\mkern-2mu M} \geq \dots \geq t_{0}$.
Note that the product and integrals in \er{Z_dyson} are omitted for the case of $M = 0$.

Each path in \er{Z_dyson} can be interpreted as a classical stochastic trajectory in continuous time with time extent $\beta$,
where the operator $\hat{K}_{x_{m-1}\to x_{m}}$ makes the instantaneous transition $x_{m-1} \to x_{m}$ at the time $t_{m}$.
We denote a stochastic trajectory of time extent $\beta$ with $M$ jumps as
\begin{equation}
    {\bm x} = \{(t_{0}, x_{0}), \dots, (t_{\mkern-1mu M}, x_{\mkern-1.4mu M})\}.
\end{equation}
Note that the last jump occurs at $t_M \leq \beta$.
We then represent \er{Z_dyson} as a sum over these jumps,
\begin{equation}
    Z_{\beta} = \sum_{\{{\bm x}\}} \delta[{x(0), x(\beta)}] \, e^{-\int_{0}^{\beta} dt \, H^{\rm c}_{x(t)}}
    \prod_{m=1}^{ \mathcal{K} \mkern-1mu({\mkern-0.4mu \bm x \mkern-0.4mu})} K_{x_{m-1}\to x_{m}} , 
    \label{Z_stochastic}
\end{equation}
where ${x}(t)$ denotes the configuration of the trajectory ${\bm x}$ at time $t$, the delta function $\delta[{x(0), x(\beta)}]$ allows only for trajectories which are periodic in time, and $\mathcal{K}({\bm x})$ is the {\em trajectory observable} (which will henceforth be denoted by a calligraphic font) which returns the number of transitions, $M$, which occur in the trajectory ${\bm x}$. 
The sum over $\{{\bm x}\}$ indicates a sum over all paths for any number of jumps, $M \in [0, \infty)$.
We can then write the probability of the trajectory ${\bm x}$ as
\begin{equation}
    P_{\beta}({\bm x}) = \frac{1}{Z_{\beta}}
    \delta[{x(0), x(\beta)}]
    e^{-\int_{0}^{\beta} dt \, H^{\rm c}_{{x}(t)}}
    \prod_{m=1}^{\mathcal{K}\mkern-1mu({\mkern-0.4mu \bm x \mkern-0.4mu})} K_{x_{m-1}\to x_{m}}.
    \label{prob_traj}
\end{equation}

The probability \er{prob_traj} can be related to that for trajectories generated by a continuous-time Markov dynamics in the following way.
We define the Markov generator $\hat{W} = \hat{K} - \hat{R}$, with the transitions given by $\hat{K}$ and their associated escape rates, $\hat{R}$, 
\beq
    \hat{R} = \sum_{x, y\neq x} K_{x\to y} \ket{x}\bra{x} = \sum_{x} R_{x} \ket{x}\bra{x}.
\eeq
Since we consider quantum systems without a sign problem \cite{Loh1990, Pan2022}, we can assume that the probabilities \er{prob_traj} are positive real numbers.
With these definitions, the relation between the Hamiltonian and the stochastic generator is
\begin{equation}
    \hat{H} = -(\hat{W} + \hat{R} - \hat{H}^{\rm c}).
\end{equation}
It follows that the trajectories that define the partition function \er{Z_stochastic} are related to those generated by the stochastic dynamics of $\hat{W}$ with the following properties:
\begin{enumerate}[(i)]
    \item The trajectory must start and finish in the 
    same configuration, i.e., it should be a {\em stochastic bridge}. A trajectory which does not meet this criterion has zero probability of occurring.

    \item The probability of a trajectory
    that satisfies (i) is  exponentially biased 
    (with respect to the probability of occurring under $\hat{W}$)
    by the time integral of  
    $\hat{H}^{\rm c} - \hat{R}$ along the trajectory. That is, its probability is multiplied by the factor 
    $e^{-\int_{0}^{\beta} d\tau \, \left( H^{\rm c}_{{x}(t)} - R_{{x}(t)} \right)}$.
\end{enumerate}
Condition (i) is needed due to the imposition of periodic paths in \er{Z_dyson}. The bias in (ii) accounts for the difference in the diagonal parts of $\hat{W}$ and $\hat{H}$.
Many methods have been developed to efficiently sample biased dynamics as in (ii). One popular approach is transition path sampling (TPS), see e.g. Refs.~\cite{Bolhuis2002, Dellago2006, Vasiloiu2020}.
Here, we will use TPS focusing on techniques for sampling trajectory updates which are local in space, cf.\ Refs.~\cite{Krzakala2008, Mora2012,Vasiloiu2020}, and which also account for (i).

\subsection{Calculating observables}
The thermal expectation value of an observable $\hat{O}$ is given by,
\beq
    \braket{\hat{O}}_\beta = \frac{\sum_{x} \braket{x | \hat{O} \, e^{-\beta\hat{H}} | x}}{Z_{\beta}} .
\eeq
This can be rewritten as an imaginary-time average by noticing that we can arbitrarily move $\hat{O}$ through the exponential 
due to the properties of the trace,
\beq
    \braket{\hat{O}}_\beta = \frac{1}{\beta}\int_{0}^{\beta} dt \, \frac{\sum_{x} \braket{x | e^{-(\beta-t) \hat{H}} \hat{O} e^{-t \hat{H}} | x}}{Z_{\beta}}.
    \label{O}
\eeq
In order to consider how to compute \er{O} from trajectories we will deal separately with the cases when $\hat{O}$ is diagonal and off-diagonal.

For the case of a diagonal operator, 
$\hat{O}_{\rm diag} = \sum_{x} O(x) \ket{x}\bra{x}$,
\er{O} can be directly written as a sum over all trajectories
with the probability given by \er{prob_traj}. This leads to 
\beq
    \braket{\hat{O}_{\rm diag}}_\beta = \beta^{-1}\sum_{\{{\bm x}\}} P_{\beta}({\bm x}) \mathcal{O}({\bm x}),
    \label{obs_diag}
\eeq
where $\mathcal{O}({\bm x}) = \int_{0}^{\beta} dt \,  O({x}(t))$ is the time-integrated trajectory observable.
That is, we average the value of the observable over all times and over all trajectories.

For an off-diagonal operator,
$\hat{O}_{\rm off-diag} = \sum_{y \neq x} O_{x\to y} \ket{y}\bra{x}$, we can expand both exponentials in \er{O}. This expansion in the integrand of \er{O} gives a sum over all trajectories which have the transition $x \to y$ at the time $t$, summed over $y \neq x$. Compared to \er{Z_stochastic}, the jumps at time $t$ appear with factors $O_{x\to y}$ rather than $K_{x \to y}$. Thus, if we want to express $\braket{\hat{O}_{\rm off-diag}}$ as a sum over trajectories
with the probability given by \er{prob_traj},
we need to account for the change $K_{x \to y} \to O_{x\to y}$ at time $t$. 
Furthermore, since we are time averaging, we can replace the integration over time by $\mathcal{K}_{x \to y}({\bm x})$ which counts the number of jumps $x\to y$ which occur in trajectory ${\bm x}$. We can then write 
\beq
    \braket{\hat{O}_{\rm off-diag}}_\beta = \beta^{-1}\sum_{\{{\bm x}\}} P_{\beta}({\bm x})
    \sum_{x \neq y}
    \frac{\mathcal{K}_{x\to y}({\bm x})}{K_{x\to y}}
    O_{x\to y}.
    \label{obs_off_diag}
\eeq

%% file: sections/optimal.tex
\section{Optimal sampling}
\label{sec: optimal}
We now explain how trajectories can be sampled efficiently and exactly from the partition function \er{Z_stochastic} by means of a rejection-free form of TPS. 
A general method for converging to an ensemble of trajectories such as \er{prob_traj} is TPS, a form of Monte Carlo sampling in trajectory space \cite{Bolhuis2002}. A typical problem with standard TPS is that acceptance of trajectory updates can become exponentially small in system size (and/or the length of the trajectory), thus slowing down convergence.  

In the language of stochastic dynamics, the procedure we now describe is sometimes referred to as obtaining the {\em Doob dynamics} \cite{Jack2010, Chetrite2015, Garrahan2016}, a proper (normalised) stochastic dynamics derived from the original generator $\hat{W}$, which generates trajectories with a conditioned/biased probability $P_{\beta}({\bm x})$, such as \er{prob_traj}.
While this approach is exact in theory, it often relies on the computation of expressions which are, in practice, analytically (and often numerically) intractable.
However, we will show later in the paper that we can in fact implement this general idea via optimal local updates.

\subsection{Edge configurations}

The first obstacle is to sample the initial/final configuration of trajectories, $x_{0}$.
While \er{Z_stochastic} has no explicit probability distribution for the initial configuration $x_{0}$, its probability will be decided by its possible transitions and its diagonal component in \er{R}.
That is, we choose some initial configuration, $x_{0}$, with probability
\beq
    P(x_{0}) = \frac{\braket{x_{0} | e^{-\beta \hat{H}} | x_{0}}}{Z_{\beta}}.
    \label{P_initial}
\eeq
Once the trajectory edges have been selected, we can then sample the remainder of the trajectory from the subset of trajectories which have the required boundary conditions.
That is, we want to sample from the reduced dynamics
\beq
    Z_{\beta}(x_{\rm i}, x_{\rm f}) = \braket{x_{\rm f} | e^{-\beta \hat{H}} | x_{\rm i}},
    \label{z_red}
\eeq
where $x_{\rm i}$ and $x_{\rm f}$ are the initial and final configurations.
Note that, although we are only interested in trajectories which have $x_{\rm i} = x_{\rm f}$, it will be useful to solve the more general problem for later for the TFIM and QTPM.

\subsection{Continuous-time dynamics} \label{sec: ctmc}

The continuous-time Monte Carlo (CTMC) method \cite{Bortz1975, Gillespie1976, Gillespie1977} is a dynamical protocol which can be employed to sample the dynamics of \er{Z_stochastic}.
While this method is straightforward for a normalised time-homogeneous dynamics, it will be useful to formulate the general case of time-dependent dynamics for what follows. 

Consider time-dependent instantaneous transition rates, $\tilde{K}_{x\to y}(t; \beta, x_{\rm f}) \geq 0$, for transitions from configuration $x$ to configuration $y$ at time $t \leq \beta$.
We use the tilde to distinguish these from the off-diagonal entries of the Hamiltonian, cf.\ \er{K}, and also allow for a dependence on the final configuration, $x_{\rm f}$, and the overall trajectory time, $\beta$.
The associated escape rate from configuration $x$ at time $t$ is
\begin{equation}
    \tilde{R}_{x}(t; \beta, x_{\rm f}) = \sum_{y\neq x} \tilde{K}_{x\to y}(t; \beta, x_{\rm f}).
\end{equation}
The escape rate in turn determines the distribution for the waiting time $\tau$, $P_x$, for a jump out of $x$ after $t$,
\begin{multline}
    P_{x}(\tau; t, \beta, x_{\rm f}) = \tilde{R}_{x}(t+\tau; \beta, x_{\rm f}) 
    \\
    \exp\left[-\int_{t}^{t+\tau} dt' \, \tilde{R}_{x}(t'; \beta, x_{\rm f})\right].
    \label{pwait}
\end{multline}
After a waiting time has been drawn from \er{pwait}, the transition into $y$ at time $t + \tau$ is chosen with probability 
\beq
    P_{x}(y \, | \, t+\tau; \, \beta, x_{\rm f}) = \frac{\tilde{K}_{x\to y}(t+\tau; \beta, x_{\rm f})} {\sum_{z\neq x} \tilde{K}_{x\to z}(t+\tau; \beta, x_{\rm f})}.
    \label{pstate}
\eeq

\subsection{Optimal transition rates} \label{sec: rates}
Given a dynamics defined by transition rates 
$\tilde{K}_{x\to y}(t; \beta, x_{\rm f})$, the CTMC method 
generates trajectories with probability density
\begin{multline}
    \tilde{P}_{\beta}({\bm x}) = 
    \delta[x_{\rm i}, x(0)] \,\,
    e^{- \int_{0}^{\beta} dt \,\, \tilde{R}_{{x}(t)}(t; \beta, x_{\rm f})}
    \\
    \prod_{m=1}^{\mathcal{K}(\bm{x})} \tilde{K}_{x_{m-1}\to x_{m}}(t_{m}; \beta, x_{\rm f})
    \label{pi_opt}
\end{multline}
for trajectories starting from $x_{\rm i}$ (where $x_{\rm f}$ acts as a parameter in the definition of the rates).
However, we are interested in sampling trajectories from the (conditioned and tilted) distribution \er{prob_traj} where the final configuration is $x_{\rm f}$, that is,
\begin{multline}
    P_{\beta}({\bm x}) = 
    \frac{\delta[{x(\beta), x_{\rm f}}] \delta[{x(0), x_{\rm i}}]}
    {Z_{\beta}(x_{\rm i}, x_{\rm f})}
    \, \, e^{- \int_{0}^{\beta} \, dt H_{{x}(t)}^{\rm c}}
    \\
    \prod_{m=1}^{\mathcal{K}(\bm{x})} K_{x_{m-1} \to x_{m}} .
    \label{pi}
\end{multline}
In the expression above we allow (for later convenience) the final configuration to be fixed to a different value from the initial one.
As written, \er{pi} is not the distribution of trajectories generated with transition rates $K_{x_{i-1} \to x_{\rm i}}$, due to \er{pi} satisfying conditions (i) and (ii) above (the former generalised to some fixed initial and final configurations), and the need for the explicit normalisation $Z_{\beta}(x_{\rm i}, x_{\rm f})$.
Furthermore, while \er{pi_opt} appears simpler than \er{pi} due to the apparent absence of the conditioning of the boundary conditions in time, this is accounted for by the fact that the transition rates in \er{pi_opt} are {\em time-dependent}.

Our aim is to find a dynamics $\tilde{K}_{x\to y}(t; \beta, x_{\rm f})$ such that $\tilde{P_{\beta}}(\bm{x}) = P_{\beta}(\bm{x})$; that is, the optimal dynamics (or {\em Doob dynamics})
for sampling $P_{\beta}(\bm{x})$ \cite{Chetrite2015}. These optimal transition rates can be computed from the probability that the transition $x\to y$ occurs at the time $t$ under the original dynamics, conditioned by the probability that the trajectory is in configuration $x$ at time $t$,
\begin{align}
    \tilde{K}_{x\to y}(t; \beta, x_{\rm f}) &= \frac{\braket{x_{\rm f} | e^{-(\beta-t)\hat{H}} \, \hat{K}_{x\to y} e^{-t\hat{H}} | x_{\rm i}}} {\braket{x_{\rm f} | e^{-(\beta-t)\hat{H}} | x} \braket{x | e^{-t\hat{H}} | x_{\rm i}}}
    \nonumber
    \\
    &= \frac{\braket{x_{\rm f} | e^{-(\beta-t)\hat{H}} | y}}{\braket{x_{\rm f} | e^{-(\beta-t)\hat{H}} | x}} K_{x\to y}
    \nonumber
    \\
    &=  \frac{Z_{\beta-t}(y, x_{\rm f})}{Z_{\beta-t}(x, x_{\rm f})} K_{x\to y},
    \label{K_opt}
\end{align}
where in the second line we have inserted the definition of $\hat{K}_{x\to y}$, \er{K}, and in the third line we used \er{z_red}. From this, we are able to calculate the escape rate from the configuration $x$ at time $t$,
\begin{align}
    \tilde{R}_{x}(t; \beta, x_{\rm f}) &= \sum_{y}  \tilde{K}_{x\to y}(t; \beta, x_{\rm f})
    \nonumber
    \\
    &= \frac{\braket{x_{\rm f} | e^{-(\beta-t)\hat{H}} \, \hat{K} | x}} {\braket{x_{\rm f} | e^{-(\beta-t)\hat{H}} | x}}
    \nonumber
    \\
    &=H_{x}^{\rm c} - \frac{\partial}{\partial t} \ln Z_{\beta - t}(x, x_{\rm f}) ,
    \label{R_opt}
\end{align}
where we have used \er{K} to account for all possible transitions out of $x$ and the derivative of \er{z_red} to obtain the last line.
Based on the rates from \era{K_opt}{R_opt}, this time-dependent dynamics gives 
$\tilde{P}_{\beta}({\bm x}) = P_{\beta}({\bm x})$ if the distribution of initial configurations is set to be $x_{\rm i}$
\footnote{Because of this, CTMC can be used with the optimal transition rates Eqs.~\eqref{K_opt} and \eqref{R_opt} to implement a TPS method which respects detailed balance with respect to \er{Z_stochastic}: $\pi({\bm x})P({\bm x}' | {\bm x}) = \pi({\bm x})\pi({\bm x}') = \pi({\bm x}')P({\bm x} | {\bm x}')$.}.
Appendix \ref{appendix:sampling} explains how the optimal transition rates can be used with the CTMC algorithm from the previous section to sample the time-dependent dynamics.

\subsection{Example: single two-level system}
\label{two-level}

\begin{figure}[t]
    \centering
    \includegraphics[width=\linewidth]{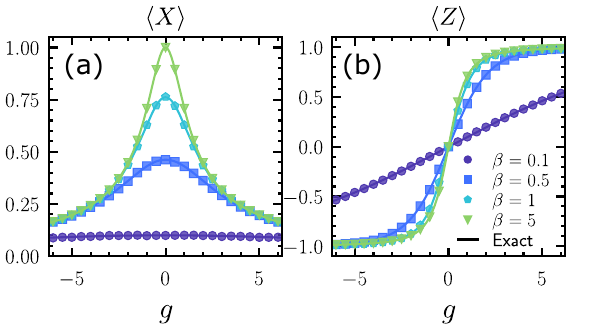}
    \caption{\textbf{Sampling of a single spin.} 
    The magnetisations (a) $\braket{\hat{X}}$ and (b) $\braket{\hat{Z}}$ for the Hamiltonian \er{Hspin}, 
    measured from trajectory sampling (data points) and compared to the exact result (solid lines) for various inverse 
    temperatures $\beta$.
    }
    \label{fig: single_spin}
\end{figure}

As an illustration of the ideas above, we consider a simple problem which will be important later in this paper: a single two-level system with the states $\sigma = \{-1, +1\}$, which we will refer to as a spin.

A generic Hamiltonian for such a system reads
\beq
    \hat{H}^{\rm spin} = -\hat{X} - g \hat{Z},
    \label{Hspin}
\eeq
where $\hat{X}$, $\hat{Y}$ and $\hat{Z}$ are the usual Pauli operators.
By calculating the matrix exponential of \er{Hspin}, is can be shown that the reduced dynamics, \er{z_red}, takes the form
\beq
    Z_{t}(\sigma_{i}, \sigma_{f}) =
    \begin{cases}
        \cosh(\theta_{t}) + \frac{g}{\sqrt{1+g^2}}\sinh(\theta_{t}), & \sigma_{i, f}=1
        \\
        \cosh(\theta_{t}) - \frac{g}{\sqrt{1+g^2}}\sinh(\theta_{t}), & \sigma_{i, f}=-1
        \\
        \frac{1}{\sqrt{1+g^2}}\sinh(\theta_{t}), &\sigma_{i} \neq \sigma_{f}
    \end{cases}
    \label{Zspin}
\eeq
where $\theta_{t} = t\sqrt{1+g^2}$.
Using \er{Zspin}, we are able to define a time-dependent stochastic dynamics which allows us to exactly sample trajectories with probabilities given by \er{prob_traj}.
The initial/final configuration $\sigma_{0}$ is chosen with probability
$
    {Z_{\beta}(\sigma_{0}, \sigma_{0})}/{\sum_{\sigma} Z_{\beta}(\sigma, \sigma)}
$,
and the time-dependent transition rates for the dynamics are
\beq
    \tilde{K}_{\sigma \to -\sigma} (t; \beta, \sigma_{0}) = \frac{Z_{\beta - t}(-\sigma, \sigma_{0})}{Z_{\beta - t}(\sigma, \sigma_{0})}.
\eeq

By simply running this dynamics we sample the trajectories we need for computing thermal averages in this problem.
The results are shown in Figs.~\ref{fig: single_spin}(a, b) for the average magnetisations $\braket{\hat{X}}$ and $\braket{\hat{Z}}$ for various inverse temperatures $\beta$.
The data points show the trajectory-averaged values using \era{obs_diag}{obs_off_diag}.
The numerical results coincide with the exact averages, 
\begin{align}
    \braket{\hat{Z}} =
    g \braket{\hat{X}} =
     \frac{g}{\sqrt{1 + g^2}} \tanh(\beta\sqrt{1 + g^2}) .
\end{align}

%% file: sections/ising.tex
\section{Transverse field Ising model in dimensions one and two}
\label{sec:ising}

We now show how the trajectory sampling approach can be used to sample from the Boltzmann distribution of the transverse field Ising model (TFIM) with $N$ spins 
\begin{align}
    \hat{H} &= -h\sum_{i=1}^N \hat{X}_i - J\sum_{\left<i, j\right>} \hat{Z}_i \hat{Z}_j
    \label{H_ising}
\end{align}
where $\hat{X}_i$
and $\hat{Z}_i$
are the Pauli operators acting on site $i$, and nearest neighbours are denoted by
$\left<i, j\right>$.
The method we use in this section can be equally adapted to accommodate for any other potential (the second summation) with the same single-body kinetic term, see App.~\ref{appendix:statistics} for an example with a global diagonal operator.
In the next section we demonstrate how to deal with other kinetic terms. 

In what follows we set $h = 1$ and consider periodic boundary conditions in space.
Sampling directly from the full partition function \er{Z_dyson} is difficult.
We instead use a single-spin update scheme, whereby we update the entire trajectory for a single spin, keeping the trajectories of all other spins fixed by sampling directly from a reduced partition function.

\subsection{Redrawing trajectories for an individual spin}
\label{sec: ising_sampling}

We now explain how our formulation can be used to perform single spin updates of the many-body trajectories. 
These updates adjust the entire trajectory of a single spin within the many-body trajectory, modelling the other spins as an effective time-dependent {\em environment} (or {\em heat bath} \cite{Krzakala2008}).
We stress that this update is equivalent to that presented in Ref.~\cite{Krzakala2008}, which considered \er{H_ising} on the Bethe lattice, and similarly that of Ref.~\cite{Mora2012} for the classical Ising model.
Here, we instead use this approach on the TFIM in 1D and 2D to demonstrate how the update can be used to investigate the ground state properties and phase transitions of quantum lattice models.

\subsubsection{Factorisation of the partition function}
We consider the many-body trajectory ${\bm \omega}$ as a collection of the $N$ individual spin trajectories, ${\bm \omega} = \{{\bm \sigma^{1}}, \cdots, {\bm \sigma^{N}}\}$, with ${\bm \sigma^{j}}$ denoting the time series of transitions for spin $j$
\beq
    {\bm \sigma^{j}} = \{ (t^{j}_{0}, \sigma^{j}_{0}), (t^{j}_{1}, \sigma^{j}_{1}), \cdots, (t^{j}_{M_{j}}, \sigma^{j}_{M_{j}})\},
    \label{spin_traj_def}
\eeq
each with a total of $M_{j}$ transitions.
The partition function becomes
\begin{multline}
    Z^{\rm TFIM}_{\beta} = 
    \sum_{\{{\bm \omega}\}}
    \left[\prod_{i=1}^{N} \delta[\sigma^i(0), \sigma^i(\beta)] \right] 
    e^{\int_{0}^{\beta} dt \, \sum_{\braket{i, j}} J {\sigma}^{i}(t){\sigma}^{j}(t)}
    \label{Z_ising}
    \\
    \prod_{j=1}^{N}
    \prod_{m=1}^{\mathcal{K}\mkern-1mu ({\bm \sigma}^j)} 
    K_{\sigma^{j}_{m-1} \to \sigma^{j}_{m}} .
\end{multline}
In the above, all $K_{\sigma \to \sigma'} = 1$, cf.\ \er{H_ising}, but we keep them in the expression to keep track of the spin transitions.

At this point, we notice that we are able to factorise \er{Z_ising} into a factor which depends on a specific ${\bm \sigma^{l}}$, and a factor which does not,
\begin{widetext}
\begin{multline}
    Z^{\rm TFIM}_{\beta} = 
    \sum_{\{{\bm \omega}\}}  
    \Bigg(
        \delta[\sigma^l \mkern-1mu (0), \sigma^l \mkern-1mu (\beta)]
        \, 
        e^{\int_{0}^{\beta} dt \, \sum_{i; \braket{i, l}} J {\sigma}^{i} \mkern-1mu  (t) \mkern+1.5mu {\sigma}^{l} \mkern-1.5mu (t)}
        \prod_{m=1}^{\mathcal{K} \mkern-1mu  ({\bm \sigma}^{l})}
        K_{\sigma^{l}_{m-1} \to \sigma^{l}_{m}}
    \Bigg)
    \\
    \prod_{p \neq l}
        \Bigg(
        \delta[\sigma^p \mkern-1mu (0), \sigma^p \mkern-1mu (\beta)]
        \, 
        e^{\int_{0}^{\beta} dt \, 
        \frac{1}{2} \sum_{j \neq l; \braket{j, p}}
        J {\sigma}^{p} \mkern-1mu (t) \mkern+1.5mu  {\sigma}^{j} \mkern-1mu (t)}
        \prod_{n=1}^{\mathcal{K} \mkern-1mu  ({\bm \sigma}^{p})} 
        K_{\sigma^{p}_{n-1} \to \sigma^{p}_{n}}
        \Biggl)
         ,
    \label{Z_ising_fact}
\end{multline}
\end{widetext}
where $\{i; \braket{i, l}\}$ denotes sites $i$ which are nearest neighbours of $l$, and $\{j \neq l; \braket{j, p}\}$ sites $j$ different from $l$ which are nearest neighbours of $p$.

We can now express \er{Z_ising_fact} as a sum over each spin $l$
and for each $l$ over the {\em partial trajectories}
\begin{equation}
    {\bm \sigma}^{(l)} = \{{\bm \sigma^{1}}, \cdots, {\bm \sigma^{l-1}}, {\bm \sigma^{l+1}}, \cdots {\bm \sigma^{N}}\},
\end{equation}
that is, the individual trajectories of all the spins other than $l$.
The first factor in \er{Z_ising_fact} depends on the trajectory of spin $l$ and its neighbouring spins through the interactions $\sum_{i; \braket{i, l}} J\sigma^{i}(t)\sigma^{l}(t)$, which we rewrite in the form of an effective time-dependent longitudinal field 
\begin{equation}
    g^{l}(t) := g(t \,|\, {\bm \sigma}^{(l)}) = \sum_{i; \braket{i, l}} J\sigma^{i}(t).
\end{equation}
The second factor in \er{Z_ising_fact} only depends on the trajectories of the spins $p \neq l$, and we denote it $\mathcal{N}({\bm \sigma}^{(l)})$.
The partition function can now be written as
\begin{widetext}
\beq
%\begin{multline}
    \label{Z_ising_factorized}
    Z^{\rm TFIM}_{\beta} = N^{-1} 
    \sum_{l} 
        \sum_{\{{\bm \sigma}^{(l)}\}}
            \mathcal{N}({\bm \sigma}^{(l)}) \,
        \sum_{\{{\bm \sigma}^{l}\}} 
            \delta[{\sigma^{l}} \mkern-1mu (0),{\sigma^{l}} \mkern-1mu (\beta)]
            \,
            e^{\int_{0}^{\beta} dt \, g^{l} \mkern-1mu (t) \mkern+1.5mu {\sigma}^{l} \mkern-1mu (t)} \,
            \prod_{m=1}^{\mathcal{K} \mkern-1mu ({\bm \sigma}^{l})}
                K_{\sigma^{l}_{m-1} \to \sigma^{l}_{m}}
    %\\
    = N^{-1} \sum_{l}  \sum_{\{{\bm \sigma}^{(l)}\}} \mathcal{N}({\bm \sigma}^{(l)}) \, Z_{\beta}^{l}[g^{l}].
%\end{multline}
\eeq
\end{widetext}

Now suppose we have some many-body trajectory ${\bm \omega}$, and we have randomly chosen to update site $l$.
The partition function for spin $l$ is $Z_{\beta}^{l}[g^{l}]$, with the time-dependent longitudinal field $g^{l}(t)$ which is piecewise constant, meaning it remains constant except for a discrete number of times, $\tilde{M}$, when it instantaneously transitions to some other value.
$\tilde{M}$ is the total number of times any of the neighbouring spins transition.
The partition function can now be written in the following way:
\beq
    Z^{l}_{\beta}[g^{l}] = \sum_{\sigma_{0}, \cdots, \sigma_{\tilde{M}}} \prod_{m=0}^{\tilde{M}} Z_{\Delta \tau_{m}}(\sigma_{m}, \sigma_{m+1}; \, g^{l}(\tau_{m})).
    \label{Z_ising_red2}
\eeq
The spin configurations $\sigma_{m} = \sigma^{l}(\tau_{m})$ are the configuration of spin $l$ at the times $\tau_{m}$ where the field $g^{l}(t)$ instantaneously changes, and we set $\sigma_{\tilde{M}+1} = \sigma_{0}$.
\Er{Z_ising_red2} is a sum over {\em all} possible spin configurations at each of these times.
The weight for each sequence of spin configurations $\{\sigma_{m}\}$ is given by the product in \er{Z_ising_red2}.
The factors $Z_{\Delta \tau_{m}}(\sigma_{m}, \sigma_{m+1}; \, g^{l}(\tau_{m}))$ are exactly the partition function of the single spin problem, \er{Zspin}, with time extent $\Delta \tau_{m} =  \tau_{m+1} - \tau_{m}$ and longitudinal field $g^{l}(\tau_{m})$, which is taken to be the value {\em after} the $m$th transition of neighbouring spins, and can be written as
\begin{widetext}
\beq
    Z_{\Delta \tau_{m}}(\sigma_{m}, \sigma_{m+1}; \, g^{l}(\tau_{m})) =
    \sum_{\{\bm{\sigma}^{l}_{m}\}} \delta[\sigma^{l}(\tau_{m}), \sigma_{m}] \, 
    \, \delta[\sigma^{l}(\tau_{m+1}), \sigma_{m+1}] 
    \, e^{g^{l}(\tau_{m}) \int_{\tau_m}^{\tau_{m+1}} dt \, \sigma^{l}(t)}
    \prod_{k_{m}=1}^{\mathcal{K}({\bm \sigma}^{l}_{m})} K_{\sigma^{l}_{k_{m}-1} \to \sigma^{l}_{k_{m}}}.
    \label{Z_ising_red3}
\eeq
\end{widetext}
In the above equation, $\bm{\sigma}^{l}_{m} = \bm{\sigma}^{l}(\tau_{m}:\tau_{m+1})$ denotes the partial trajectory of spin ${\bm \sigma}^{l}$ between times $\tau_{m}$ and $\tau_{m+1}$.
It is simple to check that \er{Z_ising_red2} with \er{Z_ising_red3} gives $Z_{\beta}^{l}[g^{l}]$.

\begin{figure*}[t]
    \centering
    \includegraphics[width=\linewidth]{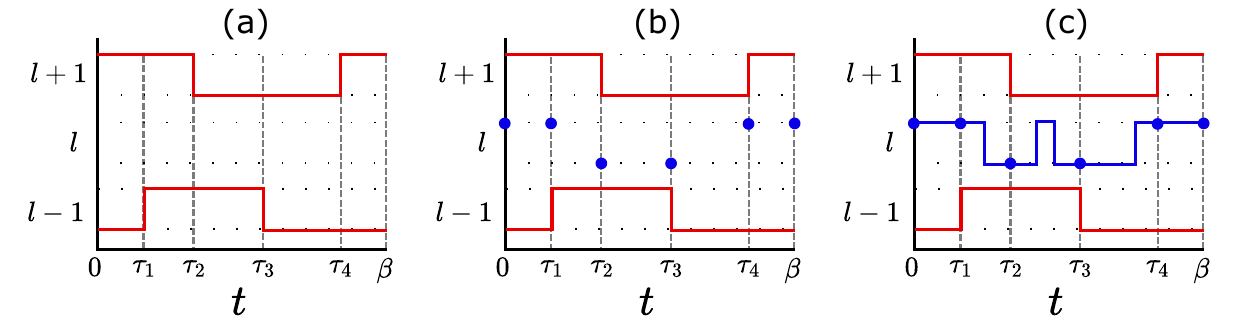}
    \caption{\textbf{Single-spin update scheme for the 1D TFIM.} 
    The spin $l$ is updated while all the other spins are kept fixed. Only the spins $l-1$ and $l+1$ (red) contribute to the dynamics of the $l$-th spin. These spins change at the times $\tau_{m}$ (dashed grey lines). The update proceeds through the following steps:
    (a) The trajectory for spin $l$ is discarded. 
    (b) We then sample the configurations $\sigma_{m}$ (blue circles) at the times in which the effective longitudinal field instantaneously changes, as explained.
    (c) We then sample bridges between each of the $\sigma_{m}$ under a static longitudinal field to construct the full trajectory (blue line).
    }
    \label{fig: update}
\end{figure*}

We now explain the strategy to update a trajectory ${\bm \omega}$, making use of the above equations and Fig.~\ref{fig: update}, which demonstrates the update for the 1D TFIM.
We first randomly select a spin $l \in \{1, \dots, N \}$ to update, keeping the trajectories for all other spins fixed.
The trajectories of spins neighbouring $l$ define the time-dependent longitudinal field $g^{l}(t)$. 
The process of selecting spin $l$ and determining the time-dependent field takes us from \er{Z_ising} to \er{Z_ising_factorized}, and allows us to sample a trajectory for spin $l$ through the partition function $Z_{\beta}^{l}[g^{l}]$.
We then exploit the fact that the longitudinal field is constant in between the transition times $\tau_{m}$ for spins neighbouring $l$.
In particular, we write $Z_{\beta}^{l}[g^{l}]$ as a sum over all {\em edge configurations}, $\sigma_{m}$, which gives \er{Z_ising_red2}.
The edge configurations can be sampled using transfer matrices, which is explained in Sec.~\ref{sec: edge_configurations}, see Fig.~\ref{fig: update}(b).
Finally, we sample {\em stochastic bridges} between configurations $\sigma_{m}$ and $\sigma_{m+1}$ as explained in Sec.~\ref{sec: stochastic_bridges}, see Fig.~\ref{fig: update}(c).

\subsubsection{Sampling the edge configurations} \label{sec: edge_configurations}
Given some time-dependent longitudinal field $g^{l}(t)$, we wish to sample the states of the spin $\sigma_{m} = \sigma^{l}(\tau_m)$ at the times $\tau_{m}$ which the longitudinal field changes value.
The probability of observing the sequence of configurations $\{\sigma_{0}, \cdots, \sigma_{\tilde{M}+1}\}$ is 
\begin{multline}
    P(\sigma_{0}, \cdots, \sigma_{\tilde{M}+1}) = 
    \\
    \frac{1}{Z_{\beta}^{l}[g^{l}]}
    \prod_{m=0}^{\tilde{M}} Z_{\Delta \tau_{m}}(\sigma_{m}, \sigma_{m+1}; \, g^{l}(\tau_{m})).
    \label{P_edge}
\end{multline}
Each of the partition functions $Z_{\Delta \tau_{m}}(\sigma_{m}, \sigma_{m+1}; \, g^{l}(\tau_{m}))$ has four possible values.
Notice that if we were to think of each $\sigma_{m}$ as a spin on a periodic 1D lattice with $\tilde{M}+1$ lattice sites, then $Z_{\Delta \tau_{m}}(\sigma_{m}, \sigma_{m+1}; \, g^{l}(\tau_{m}))$ can be thought of as a {\em transfer matrix} between lattice sites $m$ and $m+1$.

To efficiently sample the sequence of configurations, we start by sampling $\sigma_{0}$.
The probability of observing some $\sigma_{0}$  is given by
\beq
    P(\sigma_{0}) = \frac{1}{Z_{\beta}^{l}[g^{l}]} \sum_{\sigma_{1}, \cdots, \sigma_{\tilde{M}}}
    \prod_{m=0}^{\tilde{M}} Z_{\Delta \tau_{m}}(\sigma_{m}, \sigma_{m+1}; \, g^{l}(\tau_{m}))
\eeq
with $\sigma_{\tilde{M}+1} = \sigma_{0}$.
Notice that calculating $P(\sigma_{0})$ is equivalent to performing $\tilde{M}+1$ matrix multiplications (for $2\times 2$ matrices), and can then be calculated numerically.
The configurations $\sigma_{m}$ for $m = 1, \ldots, \tilde{M}$ can be calculated recursively using 
\begin{multline}
    P(\sigma_{m} \, | \, \{ \sigma_{0}, \cdots, \sigma_{m-1}\}) =
    \\
    \frac{\sum_{\sigma_{m+1}, \cdots, \sigma_{\tilde{M}}} \prod_{m=0}^{\tilde{M}} Z_{\Delta \tau_{m}}(\sigma_{m}, \sigma_{m+1}; \, g^{l}(\tau_{m}))}
    {P(\sigma_{m-1} \, | \, \{ \sigma_{0}, \cdots, \sigma_{m-2}\})}.
\end{multline}
The numerator is calculated as a multiplication over $\tilde{M} + 1 - m$ matrices and $m$ real numbers, and the denominator is known from the previous iteration of the calculation.
It is simple to show that sampling from each of these $\tilde{M}+1$ distributions yields a sequence of configurations sampled with probability given by \er{P_edge}.

\subsubsection{Sampling the full trajectory}\label{sec: stochastic_bridges}
After we have sampled the configurations $\sigma_{m}$,
all that is left to do is to sample from each of the partition functions
$Z_{\Delta \tau_{m}}(\sigma_{m}, \sigma_{m+1}  ; \, g^{l}(\tau_{m}))$.
That is, we must sample stochastic bridges between configurations $\sigma_{m}$ and $\sigma_{m+1}$ as was done previously for the two-level system.
This indicates that the trajectory between times $\tau_{m}$ and $\tau_{m+1}$ is sampled from a dynamics initialised in configuration $\sigma_{m}$, with transition rates 
\beq
    \tilde{K}_{\sigma^{l} \to -\sigma^{l}} (t; \Delta \tau_{m}, \sigma_{m+1}) = \frac{{Z}_{\Delta \tau_{m} - t}(-\sigma^{l}, \sigma_{m+1}   ; \, g^{l}(\tau_{m}))} {Z_{\Delta \tau_{m} - t}(\sigma^{l}, \sigma_{m+1}  ; \, g^{l}(\tau_{m}))}.
\eeq
Each partial trajectory between times $\tau_{m}$ and $\tau_{m+1}$ is sampled with probability
\begin{widetext}
\beq
    \label{P_partial}
    P({\bm \sigma}^{l}_{m} \, | \, {\bm \sigma}^{(l)}) =
    %\\
    \frac{\delta[\sigma_{m}, \sigma^{l}(\tau_{m})] \, \delta[\sigma_{m+1}, \sigma^{l}(\tau_{m+1})]} {Z_{\Delta \tau_{m}}(\sigma_{m}, \sigma_{m+1} ; \, g^{l}(\tau_{m}))} 
    %\\
    \, e^{g(\tau_{m}) \int_{\tau_{m}}^{\tau_{m+1}} dt \, \sigma^{l}(t)}
    \prod_{k_{m}=1}^{\mathcal{K}({\bm \sigma}^{l}_{m})} K_{\sigma^{l}_{k_{m}-1} \to \sigma^{l}_{k_{m}}}.
\eeq
\end{widetext}
Concatenating the partial trajectories gives the fully sampled trajectory for the spin ${\bm \sigma}^{l}$,
which is sampled with probability
\begin{multline}
    P({\bm \sigma}^{l} \,| \, {\bm \sigma}^{(l)})  = \frac{1}{Z_{\beta}[g^{l}]} \prod_{m=0}^{\tilde{M}} \Big[ Z_{\Delta \tau_{m}}(\sigma_{m}, \sigma_{m+1}; \, g^{l}(\tau_{m}))
    \\
    P({\bm \sigma}^{l}_{m} \, | \, {\bm \sigma^{(l)}}) \Big].
\end{multline}
Notice that the trajectories are defined to be piecewise continuous, and so we can drop the intermediate $\delta[\sigma_{m}, \sigma(\tau_{m})]$ terms in \er{P_partial},
\begin{multline}
    P({\bm \sigma}^{l} \, | \, {\bm \sigma}^{(l)}) = \frac{\delta[\sigma(\beta), \sigma(0)]}{Z_{\beta}[g^{l}]} e^{\int_{0}^{\beta} dt \, \sum_{i; \braket{i, l}} J \sigma^{i}(t) \sigma(t)}
    \\
    \prod_{k=1}^{\mathcal{K}({\bm \sigma})} K_{\sigma_{k-1} \to \sigma_k}.
\end{multline}

\subsubsection{Detailed balance}
For the trajectory update to be rejection free, it must obey detailed balance with respect to \er{Z_ising}.
It can be verified that this is true by considering the transition probabilities.
That is, given some trajectory ${\bm \omega}$, the probability to generate a new trajectory $\tilde{\bm \omega}$ is
\beq
    P(\tilde{\bm \omega} \, | \,  {\bm \omega}) = N^{-1} P(\tilde{\bm \sigma}^{l} \, | \, {\bm \sigma}^{(l)}),
\eeq
where the factor of $N$ comes from the fact that site $l$ is chosen randomly.
Notice that since $P({\bm \sigma}^{l} \, | \, {\bm \sigma}^{(l)}) \propto P_{\beta}({\bm \omega})$, it follows that
\beq
    \frac{P(\tilde{\bm \omega} \,| \, {\bm \omega})}{P({\bm \omega} \, | \, \tilde{\bm \omega})}
    = \frac{P(\tilde{\bm \sigma}^{l} \, | \, {\bm \sigma}^{(l)})}{P({\bm \sigma}^{l} \, | \, {\bm \sigma}^{(l)})}
    = \frac{P_\beta(\tilde{\bm \omega})}{P_\beta({\bm \omega})}
\eeq
and detailed balance is obeyed.

\begin{figure*}[t]
    \centering
    \includegraphics[width=\linewidth]{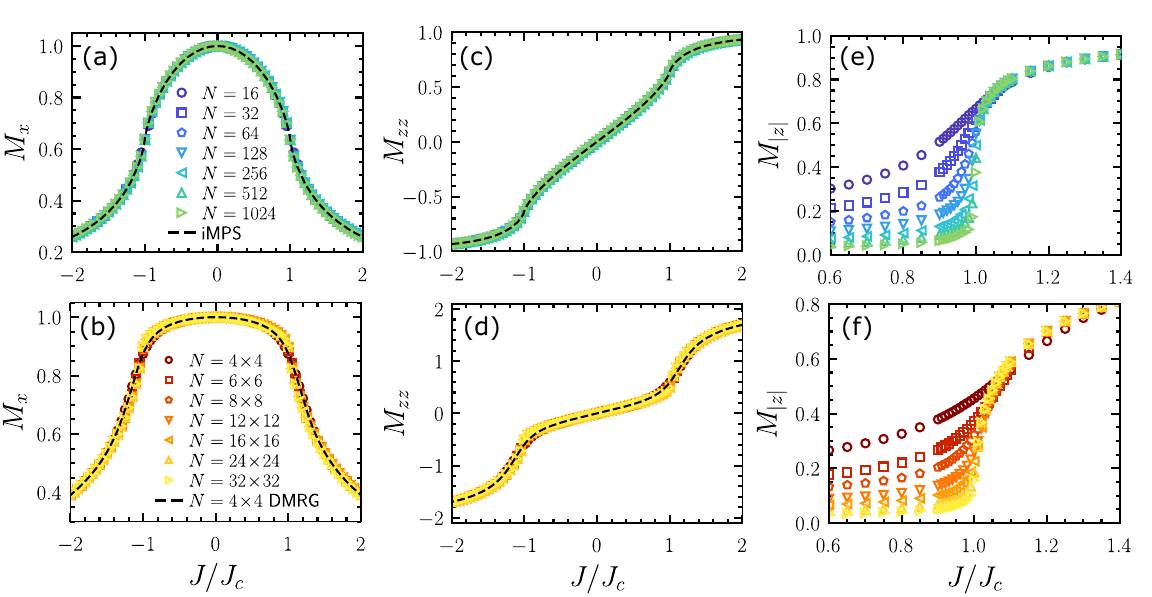}
    \caption{\textbf{Ground state properties of the TFIM.} 
    Numerical results using the single-spin update for the 1D TFIM (top) for system sizes $N = 16$ to $1024$ and the 2D TFIM (bottom) with $N = L^{2}$ and $L = 4$ to $32$.
    We show the average transverse magnetisation, $M_{x}$, in panels (a) and (b),
    the average two-spin correlator, $M_{zz}$, in panels (c) and (d) and the average magnetisation, $M_{|z|},$ in panels (e) and (f).
    Data points are the results from trajectory sampling.
    The dashed black line in the top row of panels is from the iMPS, and from the 2D-DMRG with $L = 4$ for the bottom row.
    The value of $J_{c}$ in 1D is $J_{c} = 1$ and $J_{c} \approx 0.32747$ in 2D.
    }
    \label{fig: ising_1d2d}
\end{figure*}

\subsection{Monte Carlo method}
The method of resampling a trajectory for a single spin given in the previous section can now be used to implement a Monte Carlo algorithm to sample the many-body trajectories using the following steps:
\begin{enumerate}
    \item Create some initial seed trajectory, ${\bm \omega} = ({\bm \sigma}^{1}, \cdots, {\bm \sigma}^{N})$ with inverse temperature $\beta$.
    \item Choose a random lattice site $l \in [1, N]$.
    \item Generate a new trajectory, ${\bm \sigma}^{l}$, for the spin. 
    \item Repeat from step 2 until convergence.
\end{enumerate}
There are various ways to generate an initial seed trajectory. The key requirement is periodicity in the time interval $[0, \beta]$.
The simplest way to achieve this is to generate the trajectory for each spin independently with a non-interacting dynamics, but allowing for the possibility of a local longitudinal field (see Sec.~\ref{two-level}).
The choice of the initial trajectory can be guided by the phase one is trying to target. For example, for the TFIM with $J/h > 1$ we expect to see ferromagnetic behaviour, 
and so we could choose a large local longitudinal field to force an initial trajectory with a higher likelihood of magnetic ordering. A second approach, which we discuss in Sec.~\ref{sec: tpm_thermal} is {\em thermal annealing}.

For $|J/h| > 1$, there is spontaneous breaking of the symmetry at the level of the ground state.
Indeed, depending on the initial trajectory seed, the single-spin update approach will break the $\mathbb{Z}_{2}$ symmetry and will only be ergodic over one of the ground states.
In practice, one should randomly perform the global update ${\bm \sigma}^{i}(t) \to -{\bm \sigma}^{i}(t)$ for all spins simultaneously to ensure both states are explored.
In order to investigate the effects of the local updates, we choose to not do this here.

\subsection{Continuous phase transition in the TFIM}

We now demonstrate the effectiveness of our approach for the 1D and 2D TFIM.
For fixed $h = 1$, both models are known to undergo a continuous phase transition at the critical points $J_{c} = 1$ \cite{sachdev_QPTbook} and $J_{c} \approx 0.32747$ \cite{Rieger1999, Blote2002, Huang2020}, respectively.
We investigate their ground state by using the previously described trajectory sampling algorithm with $\beta = 128$ and a variety of system sizes, $N$.
Figures~\ref{fig: ising_1d2d}(a, b) show the average transverse magnetization, 
\beq
    M_{x} = N^{-1}\sum_{i=1}^{N} \braket{\hat{X}_i},
\eeq
in 1D and 2D, respectively,
while Figs.~\ref{fig: ising_1d2d}(c, d) show the two-spin correlator,
\beq
    M_{zz} = N^{-1}\sum_{\braket{i, j}} \braket{\hat{Z}_i \hat{Z}_j}.
\eeq
For 1D, we compare our results for systems of size $N=16$ to $1024$ to those from infinite matrix product states (iMPS) \cite{Vidal2007, Orus2008}.
These methods allow us to determine the properties of the 1D chain to high accuracy.
In particular, we also use the infinite time-evolving block decimation (iTEBD) method \cite{Vidal2007, Orus2008}.
For 2D, we show results for systems $N=4 \times 4$ to $32 \times 32$.
These demonstrate agreement with the 2D density matrix renormalization group (DMRG) \cite{White1992,Schollwock2010,Stoudenmire2012} for $N = 4\times 4$ (dashed line).

The characterisation of a continuous phase transition is facilitated through the use of an order parameter which displays a discontinuous derivative at the critical point $J_{c}$ in the large system size limit.
The commonly used order parameter for Ising models is the longitudinal magnetisation,
\beq
    M_{|z|} = N^{-1} \left\langle\left|\sum_{i=1}^{N} \hat{Z}_i\right|\right\rangle.
    \label{M}
\eeq
Figures~\ref{fig: ising_1d2d}(e, f) demonstrate a sharp increase for $M_{|z|}$ at the critical point, $J / J_{c} = 1$.

\subsection{Trajectory autocorrelations}

We now investigate the autocorrelation properties of the sampling dynamics using the single-spin update for the 1D and 2D TFIM.
Under TPS we generate a {\em Markov chain of trajectories}, $\{{\bm \omega}_{0} \to {\bm \omega}_{1} \to \cdots \to {\bm \omega}_{N_{\rm traj}}\}$. In order to test the convergence to the target trajectory ensemble, cf.\ \er{prob_traj}, 
we define the autocorrelation between two trajectories in the Markov chain separated by $\mu$ TPS updates,
\beq
    \mathcal{C}({\bm \omega}_{\nu}, {\bm \omega}_{\nu+\mu}) 
    = 
    \frac{1}{\beta N}\sum_{j=1}^{N}\int_{0}^{\beta} d\tau \, 
    \sigma^j_{\nu}(\tau) \, \sigma^j_{\nu+\mu}(\tau) ,
\eeq
where $\sigma^j_{\nu}(\tau)$ indicates the value of the $j$-spin of the $\nu$-th trajectory at time $\tau$.
The trajectory ensemble average can then be estimated from the Markov chain,
\beq
    C_{\mu} 
    = 
    \frac{1}{N_{\rm traj} - \mu}
    \sum_{\nu = 1}^{N_{\rm traj}-\mu} 
    \frac{\mathcal{C}({\bm \omega}_{\nu}, {\bm \omega}_{\nu+\mu}) - \braket{\hat{Z}}^2}{1 - \braket{\hat{Z}}^2},
\eeq
where $\braket{\hat{Z}}$ is the equilibrium expectation value of $\hat{Z}_i$ (for any $i$).
With the above definition, this autocorrelator is normalized to be $C_{0} = 1$ and $C_{\infty} = 0$ for $N_{\rm traj} \to \infty$.

\begin{figure}[t]
    \centering
    \includegraphics[width=\linewidth]{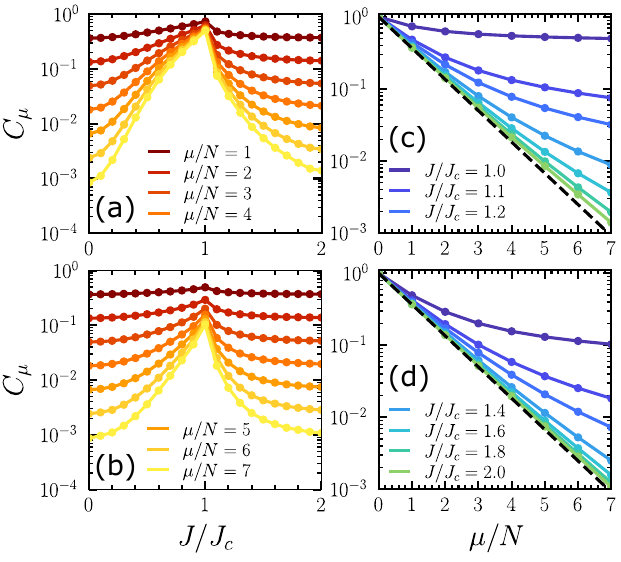}
    \caption{\textbf{Convergence in the TFIM.}
    The autocorrelation between trajectories $C_{\mu}$ as a function of $J/J_{c}$ and $\mu / N \in [1, 7]$ for
    (a) the 1D TFIM with $N = \beta = 128$ and
    (b) the 2D TFIM with $N = L^{2}$, $L = 32$ and  $\beta = 128$.
    Panels (c) and (d) show the same data but as a function of $\mu / N$ for various values of $J / J_{c}$.
    The black dashed lines show exponential decay.
    }
    \label{fig: ising_correlations}
\end{figure}

We show $C_{\mu}$ as a function of $J/J_{c}$ in Fig.~\ref{fig: ising_correlations} for $\mu / N \in [1, 7]$, for (a) the 1D TFIM and (b) the 2D TFIM.
The same data is shown in panels (c) and (d) as a function of $\mu / N$.
Far from the critical point the trajectory autocorrelation function decays approximately exponentially.
Near the critical point, this decay indicates that trajectories remain correlated even after a considerable number of TPS iterations.
This is a manifestation of the expected slow down of Monte Carlo sampling near criticality at the level of trajectory sampling.
This is a short-coming of the method when compared to methods which use non-local updates, such as rejection-free cluster updates \cite{Rieger1999, Blote2002} and Worm algorithms \cite{Beard1996, Prokofev1998, Huang2020}.
However, it is important to note that it is not always easy to find cluster updates for interaction terms which are more complex than the Ising interaction in \er{H_ising}.
On the contrary, the approach described here can be generalized to investigate arbitrary diagonal terms with a transverse field, see Ref.~\cite{Sfairopoulos2023} for triangular plaquette interactions and App.~\ref{appendix:statistics} for an example with a global diagonal operator.

%% file: sections/tpm.tex
\section{Quantum triangular plaquette model}
\label{sec:tpm}

We now show how the local update scheme can be generalised to models with more complex kinetic terms in their Hamiltonians.
As a concrete example we consider the quantum triangular plaquette model (QTPM) \cite{Yoshida2014, Devakul2019, Devakul2019b, Vasiloiu2020, Zhou2021, Sfairopoulos2023, Wiedmann2023}, a generalisation of the classical TPM studied in the context of glassy systems
\cite{Newman1999, Garrahan2000, Garrahan2002}.
This is a model defined on a triangular lattice with interactions between a subset of the triangular plaquettes
and a magnetic field transverse to them. We write the Hamiltonian of the QTPM as 
\beq
    \hat{H} = -h \sum_{\{i, j, k\} \in\tri} \hat{X}_i \hat{X}_j \hat{X}_k - J \sum_{i=1}^{N} \hat{Z}_i ,
    \label{H_tpm}
\eeq
where $\triangle$ indicates the upward pointing triangular plaquettes in the triangular lattice, see Fig.~\ref{fig: tpm_description}. We have chosen a representation of the QTPM where the interactions are off-diagonal in the classical basis and the magnetic field is longitudinal. This corresponds to the {\em dual} model of the usual QTPM.

Most numerical studies \cite{Yoshida2014, Vasiloiu2020, Sfairopoulos2023} indicate that the QTPM has a first-order quantum phase transition at $|J_{c} / h_{c}| = 1$ between two distinct phases.
In our recent paper, Ref.\ \cite{Sfairopoulos2023},
we used the method presented in Sec.~\ref{sec:ising} with single-spin updates on the Hamiltonian dual to \er{H_tpm}.
The aim of this section is to demonstrate how the update scheme can be adjusted to account for Hamiltonians like \er{H_tpm} using local updates which redraw the trajectories for multiple spins, rather than a single spin.

\subsection{Plaquette updates}

% \begin{figure*}[t]
%     \centering
%     \includegraphics[width=\linewidth]{figures/tpm_update.pdf}
%     \caption{\textbf{Plaquette updates.}
%     The update process for a triplet of spins, $i$, $j$ and $k$.
%     (a) We must first determine which other triplets each spin is a part of, and determine their flips, as shown by the red lines at times $\tau_{i}$.
%     (b) We then sample the configuration for each spin at all times $\tau_{i}$, as described in the text. This is shown by the blue dots, and the green dots show the configuration the spin transitions into if a jump happens at that time.
%     (c) The full trajectories are drawn. Notice black vertical lines illustrate jumps in the plaquette, where each spin transitions simultaneously.
%     \lc{Not happy with this figure...}
%     }
%     \label{fig: tpm_update}
% \end{figure*} 

\begin{figure}
    \centering
    \includegraphics[width=\linewidth]{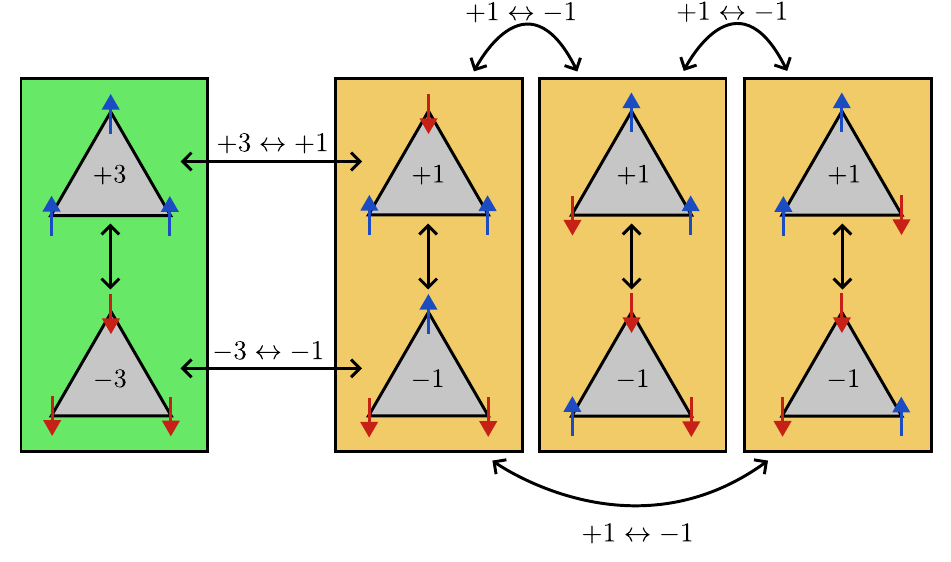}
    \caption{\textbf{Elementary transitions in the TPM.}
    A visualisation of the allowed transitions in the TPM.
    A plaquette of three spins (spins are shown by blue / red arrows), $\triangle$, can flip. 
    On the level of the individual plaquette, the plaquette can be in four different disconnected sectors, each containing two configurations and behaving like a two-level system.
    Within each sector, the system can move between the magnetisations $M_\triangle \leftrightarrow -M_\triangle$.
    The plaquettes in the green (left) box show plaquettes with total magnetisation $M_\triangle = \pm 3$.
    The three orange boxes (right) show plaquettes with total magnetisation  $M_\triangle = \pm 1$.
    The flipping of a neighbouring plaquette, $\triangle'$, causes a single spin within $\triangle$ to flip, moving $\triangle$ into a different sector, as demonstrated by the arrows connecting boxes. 
    If it moves between the $M_\triangle = |3|$ and $M_\triangle = |1|$ sectors, then the sign of the magnetisation $M_\triangle$ is preserved; otherwise, the sign is flipped.
    }
    \label{fig: tpm_description}
\end{figure}

%We now look for a local update scheme to modify the trajectories consistent with \er{Z_stochastic}.
The obvious Monte Carlo update is to randomly select a {\em plaquette} labelled by $\triangle$ encompassing three sites $\{i,j,k\} \in \triangle$.
The trajectory of this plaquette is ${\bm \eta}^\triangle = \{{\bm \sigma}^i, {\bm \sigma}^j, \bm{\sigma}^k \}$.
The corresponding transitions are those where the three spins in the plaquette flip simultaneously, in accordance with the kinetic term in \er{H_tpm}. 
Given the eight different possible states of these three spins, it might appear that one needs to consider this eight-level system in the simulations. However, as we will show below, what actually matters is the change in sign in the plaquette magnetisation,
\begin{equation}
M_\triangle = \sigma^i + \sigma^j + \sigma^k,
\end{equation}
which reduces again the analysis to that of a local two-level system, see Fig.~\ref{fig: tpm_description}. 

The time-dependent effective dynamics for the plaquette whose trajectory we choose to update is given by the six 
neighbouring plaquettes which contain any of the spins $i$, $j$ or $k$. Generalising the approach used for the TFIM above, we do not modify the trajectories of these neighbouring plaquettes when updating plaquette $\triangle$.
The neighbouring plaquettes will have a total of $\tilde{M}$ flips at times $0 \leq \tau_{1} \leq \tau_{2} \leq \cdots \leq \tau_{\tilde{M}} \leq \beta$. These transitions will force a single spin in $\eta^\triangle$ to flip
at each of these times.

\subsubsection{Factorisation of the partition function}
We start by writing the partition function as a sum over all possible imaginary-time trajectories
\footnote{This is a sum over all possible trajectories of {\em plaquettes}. Note that this is not the same as all possible trajectories of spins; the trajectories exist in a restricted space where plaquettes of spins must flip simultaneously.},
\begin{multline}
    Z^{\rm TPM}_{\beta} = 
    \sum_{\{{\bm \omega}\}} 
    e^{\sum_{i} \int_{0}^{\beta} dt \, J {\sigma}^{i} \mkern-1mu (t)} 
    \prod_{i=1}^{N} 
    \delta[\sigma^{i}(0), \sigma^{i}(\beta)]
    \\
    \prod_{\triangle}
    \prod_{m=1}^{\mathcal{K}\left({\bm \eta}^\triangle\right)}
    K_{\eta^{\triangle}_{m-1} \to \eta^{\triangle}_m},
    \label{Ztpm}
\end{multline}
where a plaquette trajectory is given by
\begin{equation}
    {\bm \eta}^{\triangle} 
    = \{ 
        (t^{\triangle}_0, \eta^{\triangle}_0), 
        (t^{\triangle}_1, \eta^{\triangle}_1), 
        \cdots, 
        (t^{\triangle}_M, \eta^{\triangle}_M)
    \} ,
\end{equation}
and 
$K_{\eta^{\triangle}_{m-1}
\to 
\eta^{\triangle}_m}$ denotes the corresponding plaquette flip.

As in the case of the TFIM, we can write \er{Ztpm} in a factorised form. Here, however, there will be three contributions. The first will contain all degrees of freedom and transitions within the plaquette $\triangle$.
The second will contain all the transitions associated with the six other plaquettes that contain any of the spins $\{i,j,k\} \in \triangle$, and it will determine the effective dynamics for the plaquette $\triangle$. 
The third factor contains all other contributions which do not affect the dynamics of plaquette $\triangle$. The partition function reads
\begin{widetext}
\begin{multline}
    Z_{\beta}^{\rm TPM} = 
        \sum_{\{{\bm \omega}\}} 
        \left[
        \prod_{i\in \triangle}\left(
        \delta[\sigma^{i}(0), \sigma^{i}(\beta)] \,
        e^{\int_{0}^{\beta} dt \, J {\sigma}^{i}(t)}\right) 
        \prod_{m=1}^{\mathcal{K}\left({\bm \eta}^\triangle\right)}
        K_{\eta^{\triangle}_{m-1} \to \eta^{\triangle}_m}
        \right] 
        %\times
        \left[
        \prod_{{\triangle'}}
        \prod_{m'=1}^{\mathcal{K}\left({\bm \eta}^{\triangle'}\right)} K_{\eta^{{\triangle'}}_{m'-1}
        \to 
        \eta^{{\triangle'}}_{m'}}
        \right]
    \\
    %\times
    \left[
        \prod_{i\not\in \triangle}\left(
            \delta[\sigma^{i}(0), \sigma^{i}(\beta)] \,
            e^{\int_{0}^{\beta} dt \, J {\sigma}^{i}(t)} 
        \right)        
        \prod_{{\triangle''}}
        \prod_{m''=1}^{\mathcal{K}\left({\bm \eta}^{\triangle''}\right)}
        K_{\eta^{{\triangle''}}_{m''-1} \to \eta^{{\triangle''}}_{m''}} 
    \right],
    \label{Z_tpm_1}
\end{multline}
\end{widetext}
where ${\triangle'}$ labels the plaquettes which share a site with $\triangle$, and ${\triangle''}$ labels those which do not.
We will now single out and propose a trajectory update for just one of the $N$ plaquettes.
The last factor does not depend on any of the spins within plaquette $\triangle$: as before, we denote ${\bm \sigma}^{(\triangle)}$ the trajectory of all the spins except from those $\{i,j,k\} \in \triangle$, and write this factor as 
$\mathcal{N}_{\triangle}\left({\bm \sigma}^{(\triangle)}\right)$.
The second factor does depend on the spins within plaquette $\triangle$, but corresponds to plaquette flips which are not for plaquette $\triangle$.
By this we mean, each lattice site $\{i, j, k\} \in \triangle$ will belong to two other plaquettes.
When either of these plaquettes flip, so will the spin $\sigma^{\{i, j, k\}}$.
However, the trajectory of each of these plaquettes is fixed, and thus we call this factor $\mathcal{M}_{\triangle}(\{ {\bm \eta}^{\triangle '} \})$, where $\{ {\bm \eta}^{\triangle '} \}$ denotes the trajectories of all neighbouring plaquettes.
The first factor corresponds to plaquette $\triangle$ and depends on $\{ {\bm \eta}^{\triangle '} \}$.
We can now rewrite $Z^{\rm TPM}_\beta$ as a sum over all plaquettes $\triangle$ and over all trajectories of plaquettes $\{{\bm \eta}^{(\triangle)}\}$ which are not $\triangle$,
\begin{multline}
    Z_{\beta}^{\rm TPM} = 
        N^{-1}
        \sum_{{\triangle}}
        \sum_{\{{\bm \eta}^{(\triangle)}\}}
        \mathcal{N}_{\triangle}\left({\bm \sigma}^{(\triangle)}\right)
        \\
        \mathcal{M}_{\triangle}(\{ {\bm \eta}^{\triangle '} \})
        Z_{\beta}^{\triangle}\left[ \{ {\bm \eta}^{\triangle '} \} \right].
\end{multline}
The partition function $Z_{\beta}^{\triangle}\left[ \{ {\bm \eta}^{\triangle '} \} \right]$ is the sum over all possible trajectories for the plaquette $\triangle$, subject to the transitions which occur in the neighbouring plaquettes at fixed times.

As was done for the TFIM, we can now factorise the partition function into $\tilde{M}+1$ components, where $\tilde{M}$ is the total number of times any of the neighbouring plaquettes flip. We can write
\beq
    Z_{\beta}^{\triangle}\left[ \{ {\bm \eta}^{\triangle '} \} \right]
    = \sum_{\eta_{0}, \cdots, \eta_{\tilde{M}}} \prod_{m=0}^{\tilde{M}} Z^{\triangle}_{\Delta \tau_{m}}(\eta_{m}', \eta_{m+1}),
    \label{Z_tpm_red}
\eeq
where $\Delta\tau_{m} = \tau_{m+1} - \tau_{m}$.
The plaquette configuration $\eta_{m}'$ is related to $\eta_{m}$ by a single spin flip, which is predetermined by the neighbouring plaquettes.
This is true for all $m \neq 0 $: periodic boundary conditions in imaginary time gives the condition $\eta_{0}' = \eta_{0} = \eta_{\tilde{M}+1}$.

\subsubsection{Determining the trajectory sector and sampling the edge configurations}

An important point to notice is that the trajectory space of ${\bm \eta}^{\triangle}$, for some chosen $\triangle$, is composed of four disconnected sectors, which we refer to as the {\em trajectory sectors}.
An initial plaquette configuration $\eta^{\triangle}_{0} = \{\sigma^{i}_{0}, \sigma^{j}_{0}, \sigma^{k}_{0}\}$ will belong to one of the four local sectors shown in Fig.~\ref{fig: tpm_description}.
At the times $\tau_{m}$ when there is a transition in a neighbouring plaquette ${\triangle'}$, the sector of the plaquette $\triangle$ will transition to one of the other three sectors, as shown in Fig.~\ref{fig: tpm_description}; however, the sector it moves to is entirely determined by which plaquette ${\triangle'}$ transitioned. Thus, the initial sector for $\eta_{0}$ will predetermine which local sector the plaquette will occupy at any time.

This observation implies that the local plaquette effectively behaves as a two-level (rather than an eight-level) system, and the flipping of neighbouring plaquettes at times $t_{m}$ changes the energies of the two levels according to the rules shown in Fig.~\ref{fig: tpm_description}.
The configurations $\eta_{m}$ in \er{Z_tpm_red} can be sampled using the transfer matrix approach as was done for the TFIM, see Sec~\ref{sec: ising_sampling}.
While the transfer matrices for the QTPM are $8\times 8$ matrices, the previous observation allows them to be treated as four separate $2 \times 2$ matrices.
This allows us to make use of the results in Sec.~\ref{sec: single_spin}, which is computationally easier than solving the original eight-level system.
To determine the trajectory sector the plaquette will lie in, we must first calculate \er{Z_tpm_red} for all four sectors, and use this as a weighting to randomly select one of the four sectors.
Then, we can sample the edge configurations $\eta_{m}$ {\em after} each time $\tau_{m}$, as was done in Sec.~\ref{sec:ising}.
This gives a sequence of configurations $\{\eta_{0}, \cdots, \eta_{\tilde{M}}\}$ which are sampled with probability
\beq
    P(\eta_{0}, \cdots, \eta_{\tilde{M}}) = \frac{1}
    {Z_{\beta}^{\triangle}\left[ \{ {\bm \eta}^{\triangle '} \} \right]}
    \prod_{m=0}^{\tilde{M}} Z^{\triangle}_{\Delta \tau_{m}}(\eta_{m}', \eta_{m+1}).
\eeq

\subsubsection{Mapping to a spin and sampling stochastic bridges}
The partition function $Z^{\triangle}_{\Delta \tau_{m}}(\eta_{m}', \eta_{m+1})$ describes the ensemble of all possible trajectories for the plaquette $\triangle$ between the times $\tau_{m} < t < \tau_{m+1}$.
Within these times, the plaquette can only hop between two distinct configurations for some chosen $\eta_{m}'$.
These configurations will have opposite magnetisation (c.f. Fig.~\ref{fig: tpm_description}).
As such, it is convenient to treat the plaquette as a spin $\kappa^{\triangle}$, with 
\beq
    \kappa^{\triangle}(t) = {\rm sgn}(\sigma^{i}(t) + \sigma^{j}(t) + \sigma^{k}(t))
\eeq
and an effective longitudinal field 
\beq
    g(t) = J|\sigma^{i}(t) + \sigma^{j}(t) + \sigma^{k}(t)|,
\eeq
which is constant between times $\tau_{m} < t < \tau_{m+1}$. We can now write
\begin{widetext}
\begin{align}
    \nonumber
    Z^{\triangle}_{\Delta \tau_{m}}(\eta_{m}', \eta_{m+1})
    &= \sum_{\{{\bm \eta}^{\triangle}_{m}\}}
    \delta[\eta_{m}', \eta^{\triangle}(\tau_{m})] \, \delta[\eta_{m+1}, \eta^{\triangle}(\tau_{m+1})]
    \, e^{J \int_{\tau_{m}}^{\tau_{m+1}} dt \, \sum_{i \in \triangle} \sigma^{i}(t)}
    \prod_{k_{m}=1}^{\mathcal{K}({\bm \eta}^{\triangle}_{m})} K_{\eta^{\triangle}_{k_{m}-1} \to \eta^{\triangle}_{k_{m}}}
    \\
    \nonumber
    &= \sum_{\{{\bm \kappa}^{\triangle}_{m}\}}
    \delta[\kappa_{m}', \kappa^{\triangle}(\tau_{m})] \, \delta[\kappa_{m+1}, \kappa^{\triangle}(\tau_{m+1})]
    \, e^{g(\tau_m) \int_{\tau_{m}}^{\tau_{m+1}} dt \, \kappa^{\triangle}(t)}
    \prod_{k_{m}=1}^{\mathcal{K}({\bm \kappa}^{\triangle}_{m})} K_{\kappa^{\triangle}_{k_{m}-1} \to \kappa^{\triangle}_{k_{m}}}
    \\
    &:= Z^{\kappa}_{\Delta \tau_{m}}(\kappa_{m}', \kappa_{m+1}; g(\tau_{m})),
\end{align}
\end{widetext}
where ${\bm \eta}^{\triangle}_{m} = {\bm \eta}^{\triangle}(\tau_{m} : \tau_{m+1})$ and ${\bm \kappa}^{\triangle}_{m} = {\bm \kappa}^{\triangle}(\tau_{m} : \tau_{m+1})$ are partial trajectories between times $\tau_{m}$ and $\tau_{m+1}$ for the plaquette and spin respectively.

Now that the partition function is expressed as the dynamics of a single spin, it can again be sampled using the results from Sec.~\ref{two-level} with a longitudinal field $g(\tau_{m})$, and then mapped back onto the plaquette.
The probability of generating each stochastic bridge is
\begin{widetext}
    \beq
    P({\bm \eta}^{\triangle}_{m} \, | \, {\bm \sigma}^{(\triangle)})
     = \frac{\delta[\eta'_{m}, \eta(\tau_{m})] \, \delta[\eta_{m+1}, \eta(\tau_{m+1})]}
     {Z^{\triangle}_{\Delta \tau_{m}}(\eta_{m}', \eta_{m+1})}
     \, e^{J\sum_{i \in \triangle} \int_{\tau_{m}}^{\tau_{m+1}} dt \, \sigma^{i}(t)}
    \prod_{k_{m}=1}^{\mathcal{K}({\bm \eta}^{\triangle}_{m})} K_{\eta^{\triangle}_{k_{m}-1} \to \eta^{\triangle}_{k_{m}}}.
     \eeq
\end{widetext}
The probability of generating the full plaquette update is the product of sampling the edge configurations and then sampling stochastic bridges between them,
\begin{multline}
    P({\bm \eta}^{\triangle} \, | \, {\bm \sigma}^{(\triangle)}) =
    \frac{\delta[\eta^{\triangle}(0), \eta^{\triangle}(\beta)]}{Z_{\beta}^{\triangle}[\{{\bm \eta}^{\triangle'}\}]}
    \\
    e^{J\sum_{i\in\triangle}\int_{0}^{\beta} dt \, \sigma_{i}(t)} \prod_{k=1}^{\mathcal{K}({\bm \eta}^{\triangle})} K_{\eta^{\triangle}_{k-1}\to \eta_{k}^{\triangle}}.
\end{multline}

\subsubsection{Detailed balance}
We can now verify that the single plaquette update obeys detailed balance and is thus rejection free.
Given some trajectory ${\bm \omega}$, the probability to generate some new trajectory $\tilde{\bm \omega}$ is
\beq
    P(\tilde{\bm \omega} \, | \,  {\bm \omega}) = N^{-1} P(\tilde{\bm \eta}^{\triangle} \, | \, {\bm \sigma}^{(\triangle)}),
\eeq
where the factor $N^{-1}$ comes from the fact the plaquette $\triangle$ is chosen at random.
Since $P(\bm \eta^{\triangle} \, | \, {\bm \sigma}^{(\triangle)}) \propto P_{\beta}({\bm \omega})$, it follows that
\beq
    \frac{P(\tilde{\bm \omega} \,| \, {\bm \omega})}{P({\bm \omega} \, | \, \tilde{\bm \omega})}
    = \frac{P(\tilde{\bm \eta}^{\triangle} \, | \, {\bm \sigma}^{(\triangle)})}{P({\bm \eta}^{\triangle} \, | \, {\bm \sigma}^{(\triangle)})}
    = \frac{P_\beta(\tilde{\bm \omega})}{P_\beta({\bm \omega})}
\eeq
and thus detailed balance is obeyed.

\subsection{Thermal annealing}\label{sec: tpm_thermal}

The expected first-order phase transition at $J_c$ can slow down the convergence for $J \approx J_c$ and large inverse temperatures, $\beta$.
In particular, if the initial seed trajectory is chosen to be in one of the two phases, there is the possibility that the update procedure cannot explore the entire trajectory space, thus remaining stuck in the incorrect phase.
This is a consequence of the large barriers which need to be overcome to move between phases.
Furthermore, making assumptions a priori on which phase the trajectory should belong to for some value of $J$ could bias the results.

A common technique used in Monte Carlo sampling to overcome such metastability is {\em thermal annealing}.
In this approach, we start from a small inverse temperature, $\beta = 0.1$, and gradually increase it to the target inverse temperature, $\beta = 128$.
Our annealing schedule is to make $N$ updates to the trajectory and then increase $\beta$ by $\Delta \beta = 0.1$ for $\beta < 32$, and $\Delta \beta = 1$ for $\beta \geq 32$.
When the inverse temperature is increased, we have to modify the trajectory to account for this.
In practice, we just stretch the trajectory time by a factor of $(\beta + \delta \beta) / \beta$.
After reaching the target inverse temperature, we do $N \times 10^2$ updates, and then restart the process.
We repeat this procedure $10^{3}$ times.

We find this approach to work well.
Indeed, when sufficiently far from $J=1$, the generated trajectories (at the target dynamics) have behaviour corresponding to their correct phase with high accuracy.
However, when close to $J=1$, the trajectories can have properties which correspond to either phase (in practice, we find that the process picks just one of the phases for each run of the annealing process).
While we cannot be certain this approach guarantees the correct amount of mixing between the phases, it provides a less biased way to propose initial trajectory seeds, and still demonstrates the first-order behaviour of the transition point.

\subsection{First-order quantum phase transition of the QTPM}

\begin{figure*}
    \centering
    \includegraphics[width=\linewidth]{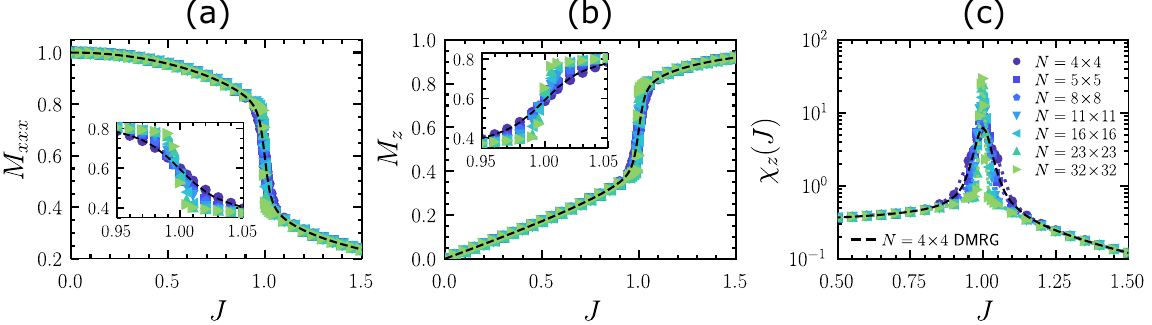}
    \caption{\textbf{Ground state phase transition in the QTPM.}
    Results using TPS with exact plaquette updates and simulated annealing for $N = L \times L$ and $L = \{4, 5, 8, 11, 16, 23, 32\}$.
    We show (a) the three spin correlator, $M_{xxx}$, (b) the longitudinal magnetization, $M_{z}$ and (c) the magnetic susceptibility, $\chi_{z}(J) = dM_{z} / dJ$.
    The data points show results from trajectory sampling, and for comparison, the dashed lines show results from DMRG for $N = 4 \times 4$.
    The insets show the behaviour close to the transition point. 
    }
    \label{fig: tpm}
\end{figure*}

We now demonstrate how the plaquette update scheme with temperature annealing can be used to investigate the quantum phase transition of the QTPM.
As described in Ref.~\cite{Sfairopoulos2023}, the finite size study of the 
QTPM has to be done with care, since, depending on the size and aspect ratio of the system used, \er{H_tpm} can have many or no symmetries. 
We focus on square lattices $N = L \times L$, with $L = \{4, 5, 8, 11, 16, 23, 32\}$  chosen such that there are no such symmetries \cite{Sfairopoulos2023}. We also compare our results against those from 2D DMRG for $N = 4 \times 4$.

Figure~\ref{fig: tpm}(a) shows the average transverse three-spin correlator,
\beq
    M_{xxx} = N^{-1}\sum_{\{i, j, k\} \in \tri} \braket{\hat{X}_i \hat{X}_j \hat{X}_k}.
\eeq
The inset shows the results close to the transition point.
Figure~\ref{fig: tpm}(b) shows the average longitudinal magnetisation,
\beq
    M_{z} = N^{-1}\sum_{i=1}^{N} \braket{\hat{Z}_i}.
\eeq
As we increase the system size, the crossover between the two phases becomes increasingly step-like, suggesting a first-order singularity in the large-size limit.
The behaviour of the magnetic susceptibility,
\beq
    \chi_{Z}(J) = \frac{dM_{z}}{dJ},
\eeq
is consistent with this interpretation, with its peak getting higher and narrower around $J = 1$ with the system size, see Fig.~\ref{fig: tpm}(c). 

%% file: sections/conclusions.tex
\section{Conclusions}
\label{sec: conclusions}

In this paper we have leveraged the connection between the continuous-time expansion of the quantum Boltzmann distribution and the rare-event sampling in stochastic dynamics for systems which have no sign problem.
In particular, we have focused on the so-called \textit{stoquastic} Hamiltonians, where 
computing the partition sum is equivalent to sampling  imaginary-time trajectories of a continuous-time Markov chain.
Each trajectory in the ensemble is conditioned to return to the initial configuration, and its probability is exponentially biased (or tilted) due to the difference between the Hamiltonian and the associated stochastic generator.
Such trajectories can be accessed with a method like TPS, i.e.\ Monte Carlo for trajectory ensembles. Specifically, we showed that in systems with finite-state local degrees of freedom (such as spins on a lattice) one can use an approach similar to that of Refs.~\cite{Krzakala2008,Mora2012} to devise a rejection-free TPS scheme, by means of an exact local generation of trajectory updates which is especially simple in spin-$1/2$ models.
In fact, as we showed above, this gives the optimal, or Doob, dynamics for sampling the rare trajectories. 

We illustrated the effectiveness of this approach by studying the quantum phase transitions of two classes of models. The first included the TFIM in 1D and 2D, where the transition is well known to be continuous. We showed that the rejection-free TPS method correctly characterises their quantum phase transition, even in the near-critical regime where trajectories take many TPS iterations to decorrelate.
The second class of models we considered are quantum plaquette models. In particular, we studied the QTPM, and showed how to generalise the rejection-free method to local multi-spin updates. While less understood than the TFIM, the QTPM has a quantum phase transition which is first-order. Again, the rejection-free method efficiently recovered the quantum phase transition.
As a by-product we also computed to high accuracy the statistics of dynamical observables (both in the large deviation regime and for finite imaginary times) in the trajectory ensembles that resolve the quantum partition sums of both the TFIM and the QTPM, which is shown in Appendix~\ref{appendix:statistics}. 

The method described here was used recently in Refs.~\cite{Sfairopoulos2023,Sfairopoulos2023b}, where the ground state phase transitions of various spin models were characterised for large system sizes.
We foresee that our approach will be useful for numerical investigations of glassy models, where non-local updates could be difficult to formulate.
While the method we used here is based on local updates, the key property is not locality but simplicity, which allows the sampling of exact trajectory moves by solving \era{K_opt}{R_opt} that define the Doob dynamics. It would be interesting to find other (perhaps non-local) moves that also solve \era{K_opt}{R_opt}. 
While finding exact updates might prove difficult, one might be able to formulate an 
approximately optimal dynamics, with the error accounted for in an acceptance test, such as Metropolis in a TPS scheme. The difficulty here is finding a dynamics which provides improvements in sampling while defined in terms of transition rates which remain computationally cheap (perhaps building on the use of tensor networks for optimal sampling \cite{Causer2022}).
A second proposition could be to implement more advanced sampling methods to overcome local minima, such as {\em parallel tempering}.
While this could present some technical challenges, it might be advantageous for the sampling of glassy models.

%% file: appendix/sampling.tex
\section{Sampling time-dependent optimal dynamics}
\label{appendix:sampling}

Suppose we are able to resolve the optimal dynamics given in Sec.~\ref{sec: rates}, with the transition rates given by \er{K_opt}.
This is a time-dependent dynamics which is defined through the reduced dynamics \er{z_red}.
To sample dynamics from it, we follow the process drawn out in Sec.~\ref{sec: ctmc}, which requires randomly drawing waiting times from the distribution \er{pwait}, and then randomly selecting a configuration to jump to from the distribution \er{pstate}.

Determining a random transition time, $\tau$, is easily done by considering the cumulative distribution function of \er{pwait},
\beq
    C_{x}(\tau; \beta, x_{\rm f}, t) = 1 - \exp\left[-\int_{t}^{t+\tau} dt' \, \tilde{R}_{x}(t'; \beta, x_{\rm f})\right].
\eeq
It is well understood that by drawing some uniformly random number $r \in [0, 1]$, and inverting $C_{x}(\tau; \beta, x_{\rm f}, t) = r$, one can pick $\tau$ with a probability density given by \er{pwait}.
Using \er{R_opt}, we find 
\beq
    C_{x}(\tau; \beta, x_{\rm f}, t) = 1 - \frac{Z_{\beta-t-\tau}(x; x_{\rm f})}{Z_{\beta-t}(x; x_{\rm f})} e^{-\tau H_{x}^{c}}.
    \label{C_z}
\eeq
While analytically inverting \er{C_z} is difficult, this can be done numerically to arbitrary accuracy using the bisection method, if one is able to calculate $Z_{t}(x; x_{\rm f})$.
This is due to the fact that $C_{x}(\tau; \beta, x_f, t)$ is a monotonically increasing function between $C_{x}(0; \beta, x_{\rm f}, t) = 0$ and $C_{x}(\beta - t; \beta, x_{\rm f}, t) < 1$. 
Note that if $r >  C_{x}(\beta - t; \beta, x_{\rm f}, t)$, then no transition time is drawn, and the system remains in state $x$ until the terminating trajectory time, $\beta$.

Once a transition time, $\tau$, has been drawn, the state which the system transitions to can be determined using \er{pstate},
\beq
    P_{x}(y \, | \, t+\tau; \beta, x_{\rm f}) = \frac{K_{x\to y} \, Z_{\beta - t - \tau} (y; x_{\rm f})} {\sum_{z\neq x} K_{x\to z} \, Z_{\beta - t - \tau} (z; x_{\rm f})}.
\eeq

%% file: appendix/lds.tex
\section{Trajectory statistics}
\label{appendix:statistics}

The connection between the quantum partition function and the ensemble of (conditioned/biased) stochastic trajectories
naturally motivates the investigation of the {\em trajectory statistics} of the imaginary-time dynamics.
Given some trajectory ensemble, such as one defined by \er{prob_traj}, and some inverse temperature (or {\em trajectory time}), $\beta$, we can define its probability distribution function over a trajectory observable, ${\cal O}({\bm x})$, through 
\beq
    P_{\beta}({\cal O}) = \sum_{\{{\bm x}\}} 
    \pi({\bm x})
    \delta \left[ {\cal O}({\bm x}) - {\cal O} \right] .
    \label{P_o}
\eeq

For simplicity, we will only consider trajectory observables which are obtained by time integrating 
diagonal operators,
\beq
    {\cal O}({\bm x}) = 
    \int_{0}^{\beta} dt \, 
    O({x}(t)).
    \label{calO}
\eeq
In practice, the computation of \er{P_o} for some arbitrary inverse temperature $\beta$ is difficult.
However, in the low temperature limit, $\beta \to \infty$, one can estimate \er{P_o} using the framework of large deviation (LD) theory (for reviews, see e.g. Refs.~\cite{Touchette2009, Garrahan2018, Jack2020, Limmer2021}). 

In the low temperature limit both the probability distribution of ${\cal O}$ and its moment generating function (MGF), 
\beq
    Z_{\beta}(s) = \sum_{\{{\bm x}\}} P_{\beta}({\bm x}) \, e^{-s {\cal O}({\bm x})},
    \label{mgf}
\eeq
take the LD forms
\begin{equation}
    P_{\beta}({\cal O}) \asymp e^{-\beta \varphi({\cal O} / \beta)} , 
    \;\;\;
    Z_{\beta}(s) \asymp e^{\beta \theta(s)},
    \label{LDs}
\end{equation}
respectively.
Here, $\varphi$ is the {\em rate function}, and $\theta$ is the {\em scaled cumulant generating function} (SCGF), with the two related through the Legendre transform, $\theta(s) = -\min_{o}\left[s{o} + \varphi(o) \right]$, where $o = \mathcal{O} / \beta$.

The MGF has a form similar to the partition function, \er{Z}. 
In terms of trajectories, it reads, cf.\ \er{Z_dyson},
\begin{align}
    Z_\beta(s) 
    &= \sum_{\{{\bm x}\}} P_{\beta}({\bm x}) e^{-s\int_{0}^{\beta} dt \, O({{x}(t))}} 
    \label{mgf2}
    \\
    &= \frac{1}{Z_{\beta}} \sum_{\{{\bm x}\}} 
    \delta[{x}(0), {x}(\beta)]
     e^{-\int_{0}^{\beta} dt \, [H^{\rm c}_{{\bm x}(t)} + s \, O({x}(t))]}
     \nonumber
    \\
    & 
    \phantom{xxxxxxx}
    \prod_{m=1}^{\mathcal{K}({\bm x})} K_{x_{m-1}\to x_{m}}.
    \label{mgf_stochastic}
\end{align}
Furthermore, from \er{mgf_stochastic} we also see that the MGF can be written as 
\beq
Z_{\beta}(s) = \frac{\Tr[e^{-\beta \hat{H}_{s}}]}{Z_{\beta}} ,
\eeq
where $\hat{H}_{s}$ is a {\em tilting} \cite{Touchette2009} of the original Hamiltonian,
\beq
    \hat{H}_{s} = \hat{H} + s \hat{O}.
    \label{Hs}
\eeq
This means that in the limit of small temperatures, \er{LDs}, we get that $\theta(s) = E_{s} - E$, where $E_{s}$ is the ground state energy of $\hat{H}_{s}$ and $E$ that of $\hat{H}_{s=0} = \hat{H}$. For the following we use this to connect quantum phase transitions to transitions in the LD statistics of trajectory observables.

\subsection{TFIM}
\begin{figure}[t]
    \centering
    \includegraphics[width=\linewidth]{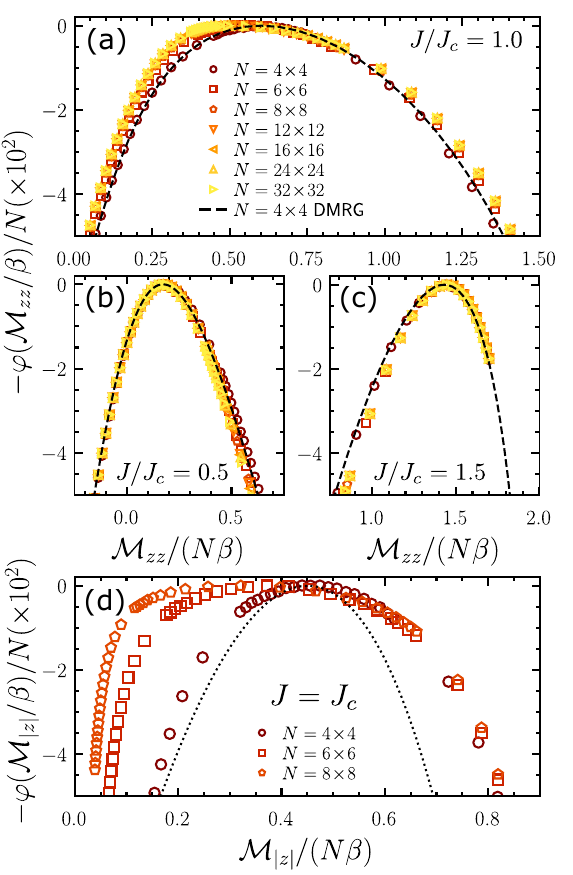}
    \caption{\textbf{Dynamical fluctuations of the imaginary-time trajectories of the 2D TFIM.}
    The low-temperature trajectory statistics are encoded in the dynamical LDs.
    We show the rate function, $-\varphi({\mathcal M}_{zz} / \beta) / N$, for the two-point correlator, $M_{zz}$, for (a) $J/J_{c} = 1.0$, (b) $J/J_{c} = 0.5$, and (c) $J/J_{c} = 1.5$.    
    The data points show results from trajectory sampling with $\beta = 128$.
    The dashed line shows the 2D DMRG for $N = 4 \times 4$.
    (d)
    Rate function $-\varphi({\mathcal M}_{|z|} / \beta) / N$ for $J/J_{c} = 1.0$ and $\beta = 64$.    
    The dotted line shows the rate function for the time-integral of the longitudinal magnetization of a single spin from \er{Hspin}, with $\braket{\hat{\sigma}^{z}} = 0.45$.
    }
    \label{fig:ising_lds}
\end{figure}

We first use the methods of Sec.~\ref{sec:ising} to investigate the trajectory statistics of the TFIM in 2D.
The first trajectory observable we consider, cf.\ \er{calO}, is the time integral of the two-point correlator,
\beq
    {\mathcal M}_{zz} ({\bm \omega}) = \int_0^\beta d\tau \sum_{\langle i, j\rangle} {\sigma}^{i}(\tau) \, {\sigma}^{j}(\tau).
\eeq 
We focus on the low-temperature limit, $\beta \gg 1$, corresponding to the ground state behaviour, where LD theory can be applied.
The LD statistics are retrieved from the ground state properties of the tilted Hamiltonian,
\beq
    \hat{H}_{s} = -h\sum_{i=1}^{N} \hat{X}_i - (J-s)\sum_{\left<i, j\right>} \hat{Z}_i \hat{Z}_j,
    \label{H_ising_s_zz}
\eeq
which is the same as the original Hamiltonian with $J \to J - s$.
Figure \ref{fig:ising_lds} shows the rate function,
$-\varphi({\mathcal M}_{zz} / \beta)$, c.f. \er{LDs},
%$
%    -\varphi({\mathcal M}_{zz}) \asymp {\log P_{\beta}(\beta {\mathcal M}_{zz})} / {\beta}
%$,
for the 2D TFIM with (a) $J/J_{c} = 1$, (b) $J / J_{c} = 0.5$ and (c) $J / J_{c} = 1.5$.
At criticality, we observe a broadening in this distribution, demonstrating the divergence in correlation lengths.
In contrast, away from the critical point the distributions become narrower.

The second trajectory quantity we consider is the time-integral of the order parameter,
\beq
    {\mathcal M}_{|z|} (\bm \omega) = \int_0^\beta d\tau \left| \sum_{i=1}^{N} {\sigma}^{i}(\tau) \right|.
\eeq
The corresponding tilted Hamiltonian is
\beq
    \hat{H}_{s} = -h\sum_{i=1}^{N} \hat{X}_i - J\sum_{\left<i, j\right>} \hat{Z}_i \hat{Z}_j + s\left|\sum_{i=1}^{N} \hat{Z}_i\right|.
    \label{H_ising_s_m}
\eeq
To simulate dynamics with \er{H_ising_s_m}, we can use the single-spin update scheme previously described, but we must now consider the trajectories of \textit{all} other spins when constructing the time-dependent dynamics, due to the coupling introduced via $M_{|z|}$ in \er{H_ising_s_m}. 
While this makes the procedure more costly, we are still able to run dynamics for moderate $N$.
We show the rate function in Fig.~\ref{fig:ising_lds}(d).
Here, the broadening at criticality is more pronounced; for comparison, we show the rate function for the longitudinal magnetisation of a single spin (dashed line).
This broadening suggests a diverging magnetic susceptibility in the large size limit.

\subsection{QTPM}

\begin{figure}[t]
    \centering
    \includegraphics[width=\linewidth]{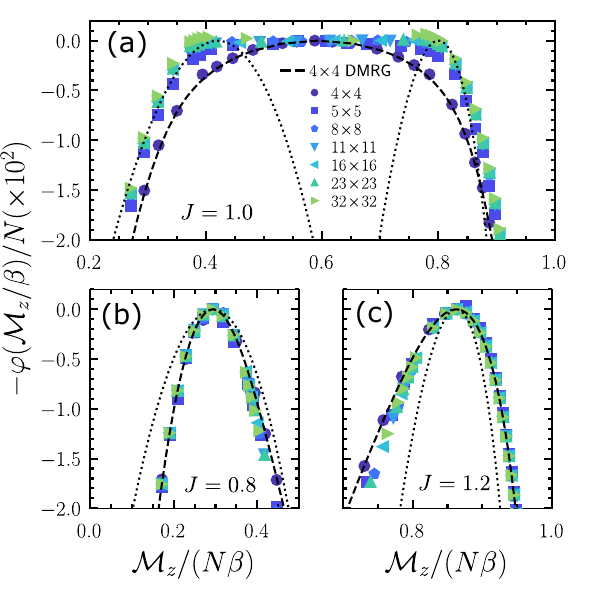}
    \caption{\textbf{Large deviations in the long imaginary-time trajectories of the QTPM.} 
    The long imaginary-time trajectory statistics can be captured through the LDs. 
    We show the rate function, $-\varphi({\mathcal M}_{z}) / N$, for (a) $J = 1.0$, (b) $ J = 0.8$ and (c) $ J = 1.2$.
    The data points show results from trajectory sampling for $N=L \times L$ with $L = \{4, 5, 8, 11, 16, 23, 32\}$, and the dashed line from 2D DMRG for $L = 4$.
    For comparison, the dotted lines show the rate functions for the single spin \er{Hspin} case, where $g$ is chosen to fit the mean to the peak of the QTPM distributions. 
    }
    \label{fig: tpm_lds}
\end{figure}

\begin{figure}[t]
    \centering
    \includegraphics[width=\linewidth]{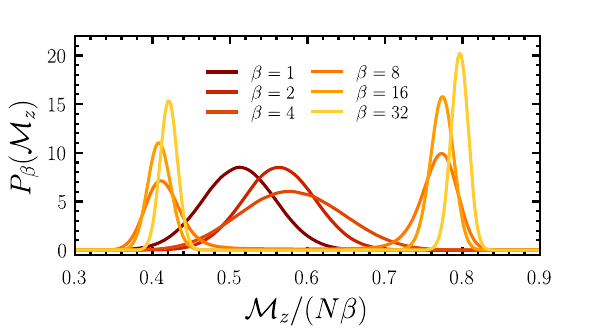}
    \caption{\textbf{Trajectory statistics at finite imaginary times in the QTPM.} 
    The estimated distribution of the time-integrated magnetisation
    for $L = 16$ for various finite times (inverse temperatures) at $J = 1$ from trajectory sampling.
    The curves are calculated using a Gaussian kernel with $\delta = 0.005$.
    }
    \label{fig: tpm_statistics}
\end{figure}

We now investigate the effects of the first-order phase transition of the QTPM at the level of the imaginary-time trajectories. We consider the statistics of the 
trajectory observable
\beq
    {\mathcal M}_{z}({\bm \omega}) = \int_0^\beta d\tau \sum_{i=1}^{N} \sigma^{i}(\tau)
\eeq
They are encoded in the tilted Hamiltonian,
\beq
    \hat{H}_{s} = -h \sum_{\{i, j, k\} \, \in \tri} \hat{X}_i \hat{X}_j \hat{X}_k - (J - s) \sum_{i=1}^{N} \hat{Z}_i,
    \label{H_tpm_s}
\eeq
which corresponds to the original Hamiltonian with $J \to J - s$.
Figure \ref{fig: tpm_lds} shows the rate function $-\varphi({\mathcal M}_{z} / \beta)$
%\beq
%    \frac{\log P_{\beta}(N\beta {\mathcal M}_{z})}{\beta} \asymp -\varphi({\mathcal M}_{z})
%    \nonumber
%\eeq
for various $J$.
The rate function at $J_c$ is shown in Fig.~\ref{fig: tpm_lds}(a): the rate function flattens with increasing system size, indicating the existence of large fluctuations. 
This behaviour is characteristic of a (dynamical) first-order phase transition due to the coexistence of two distinct (dynamical) phases. For comparison, we show the LDs of the single-spin problem from \er{Hspin} (dotted lines), where the value of $g$ is chosen to fix the mean of the distribution.
Figure \ref{fig: tpm_lds}(a) shows the LDs of the single-spin problem with $\braket{M_{z}} = 0.42$ and $0.8$, which approximately match the tails of the distribution for the QTPM.
The interpretation is clear: the two coexisting phases are homogeneous phases of distinct ${\mathcal M}_{z}$, and intra-phase fluctuations are uninteresting (thus the modes are well approximated by a single spin); the broadening in $\varphi$ in the QTPM is due to a Maxwell construct between these modes due to the fact that intermediate values of ${\mathcal M}_{z}$ are realised by coexistence, i.e.\ space-time regions of one phase separated by sharp interfaces from space-time regions of the other phase. 
In Figs.~\ref{fig: tpm_lds}(b) and (c) we also show the rate functions away from the transition point for $J = 0.8, 1.2$, respectively. While there is a slight broadening in the tails of these distributions, they still
describe single phases far from coexistence.

It is also possible to reasonably estimate the probability distribution $P_{\beta}({\mathcal M}_{z})$ for finite $\beta$ through sampling.
This is shown in Fig.~\ref{fig: tpm_statistics} at the transition point for $L = 16$, for various inverse temperatures.
We use our annealing strategy to sample $N_{\rm traj} = 10^6$ trajectories for each inverse temperature.
The results are used to approximate the probability distribution function using the Gaussian kernel,
\beq
    P_{\beta}(\mathcal{M}_{z}) \sim \sum_{i = 1}^{N_{\rm traj}} \exp\left(-\frac{[\mathcal{M}_{z} - \mathcal{M}_{z}({\bm \omega}^{i})]^2}{2\delta^2} \right),
\eeq
where $\mathcal{M}_{z}({\bm \omega}^{i})$ is the time-integrated magnetisation of the $i$-th trajectory, and $\delta = 0.005$ is the width of the filter.
Notice that at small inverse temperatures the distribution looks approximately Gaussian. However, with increasing $\beta$, the distribution becomes bimodal, demonstrating explicitly the coexistence of phases.
For the largest inverse temperatures shown here, trajectories with phase coexistence (trajectories with time-averaged $\mathcal{M}_{z}$ between the two phases) are improbable.
The broadening observed in Fig.~\ref{fig: tpm_lds}(a) can only be seen for inverse temperatures much larger than the ones considered here, where phase coexistence within trajectories can be realised by rare transitions between the two phases.